\documentclass[review,3p,times,onecolumn,numbers]{elsarticle}

\usepackage{placeins}
\usepackage{amssymb}
\usepackage{amsmath}
\usepackage{lineno}
\usepackage{gensymb}
\usepackage[lofdepth=1]{subfig}
\usepackage[table]{xcolor}
\usepackage{graphicx}
\usepackage{upgreek}
\usepackage{hyperref}
\usepackage{makecell}
\usepackage{url}
\usepackage{float}
\usepackage{stfloats} 
\usepackage{rotating}
\usepackage{afterpage}  

\journal{Nuclear Inst. and Methods in Physics Research, A}

\begin{document}

\begin{frontmatter}


\title{Mortality of ultra-thin LGADs and PiN diodes from high energy deposition}

\author[1]{A. Tishelman-Charny}
\author[2]{A. Buzzi}
\author[3]{F. Capocasa}
\author[1]{G. D'Amen\corref{cor1}}
\ead{gdamen@bnl.gov}
\author[4]{S. Diaw}
\author[2]{D. Duan}
\author[5]{M. H. Mohamed Farook}
\author[3]{\\G. Giacomini}
\author[1]{M. Kurth}
\author[2]{D. Ponman}
\author[2]{J. Roloff}
\author[3]{E. Rossi}
\author[1]{S. Stucci}
\author[1]{A. Tricoli}
\author[6]{H. Zhang}

\cortext[cor1]{Corresponding author.}

\affiliation[1]{organization={Physics Department, Brookhaven National Laboratory},
            city={Upton},
            postcode={11973}, 
            state={NY},
            country={USA}}

\affiliation[2]{organization={Physics Department, Brown University},
            city={Providence},
            postcode={02912}, 
            state={RI},
            country={USA}}

\affiliation[3]{organization={Instrumentation Department, Brookhaven National Laboratory},
            city={Upton},
            postcode={11973}, 
            state={NY},
            country={USA}}

\affiliation[4]{organization={Cheikh Anta Diop University},
            city={Dakar},
            country={Senegal}}

\affiliation[5]{organization={Department of Physics and Astronomy, University of New Mexico},
            city={Albuquerque},
            postcode={87131}, 
            state={NM},
            country={USA}}

\affiliation[6]{organization={Stuyvesant High School},
            city={New York},
            postcode={10282}, 
            state={NY},
            country={USA}}

\begin{abstract}
Low Gain Avalanche Diodes are prime candidates for high-resolution timing applications in High Energy Physics, Nuclear science, and several other fields. Operating these devices in high-radiation environments presents various hazards, including the risk of their permanent degradation or destruction caused by effects such as Single Event Burnout. Studies using minimum ionizing particles found a greatly reduced Single Event Burnout risk by operating below a bias voltage corresponding to an average electric field of 12 V/$\upmu$m - however, as high energy particle colliders produce a wide energy spectrum of radiation, it is crucial to understand this phenomenon and other possible damage mechanisms at energy deposition levels greater than those of minimum ionizing particles. This was achieved by pre-irradiating LGADs and PiN diodes with active thicknesses of 20, 30, and 50 $\upmu$m up to 1.5 $\times$ 10$^{15}$ $\mathrm{n_{eq}/cm^2}$, and exposing them to beams of protons and heavy ions (C, O, Fe, Au) at the BNL Tandem van de Graaff accelerator. Several mortality categories were observed, defined by different electrical and mechanical damage signatures. This furthers our understanding of permanent radiation damage of Silicon devices, crucial towards mitigating Single Event Burnout and other damage mechanisms to safely operate future detectors.
\end{abstract}

\end{frontmatter}


\section{Introduction}
\label{sec:introduction}

Low Gain Avalanche Diode (LGAD) \cite{pellegrini_technology_2014} sensors are a key technology for precision timing in high-energy physics, providing per-hit timing resolution of $\mathcal{O}$(10) ps \cite{lgad_review} through controlled avalanche multiplication in regions of very high electric field. Their widespread use in 4D tracking and timing layers at the upcoming High Luminosity Large Hadron Collider \cite{HGTD, MTD}, and planned use in future collider experiments \cite{IDEA, bell2025maianewdetectorconcept, ALLEGRO, ePIC} requires reliable operation in high radiation environments, where not only cumulative damage but also rare, destructive single-event effects can limit detector lifetime.

While the radiation response of LGADs to displacement damage from minimum ionizing particles (MIPs) has been extensively studied \cite{Kramberger_2015}, destructive single-event phenomena caused by non-MIP particles remain poorly characterized. In particular, Single Event Burnout (SEB) and other radiation-triggered causes of mortality represent critical failure modes in which a single ionizing particle induces abrupt and irreversible sensor breakdown. 

The first identification of SEB was in high-voltage power semiconductor devices, notably MOSFETs and bipolar transistors, as a catastrophic failure mode induced by a single ionizing particle traversing a high-field region \cite{4334670}. It is understood that SEB is triggered by thermal runaway of deposited charge in semiconductors \cite{osti_6599023}. Previous studies of heavy-ion-induced SEB in Silicon power devices exhibit a well-defined stopping power threshold for triggering SEB that decreases with increasing bias voltage and depends on ion species and energy, suggesting a possible correlation \cite{211357, Schwank1993LETThreshold, Dachs1996IonSpecies, Waskiewicz1998SOA, MAIER1998596} - simulations of heavy ion induced SEB in Silicon power devices have also been conducted \cite{1589182}. These studies establish that SEB occurs in Silicon-based power devices, and forms a basis for extending the field of study to Silicon-based detectors.

Device-level models describe SEB as originating from the localized deposition of energy along the ionizing particle's path, producing a dense electron–hole plasma that transiently perturbs the internal electric field. In regions biased near the critical field for avalanche multiplication, this perturbation may enhance this effect and may generate a highly localized current filament, ultimately driving thermal runaway and irreversible junction damage \cite{Sola2021LGAD}. This hints that beams of particles in the MIP regime may be less likely to induce the extreme charge densities, the amount of deposited charge per unit volume normalized to density, required to trigger SEB-like mechanisms, expected to be at least about 1 MeV$\boldsymbol{\cdot}$cm$^{2}$/g. In contrast, highly ionizing particles deposit orders of magnitude more energy per unit length, leading to denser carrier generation. This suggests that the probability of catastrophic failure might depend on the particle's stopping power - corresponding to the amount of energy lost by the incident particle during interactions with sensors. This dependence was systematically studied in Silicon-based sensors with a femtosecond laser \cite{Laser_Jinst, LASTOVICKAMEDIN2022167321}, revealing several findings: it was observed that SEB is not an LGAD-specific effect, as the effect was produced in both LGADs and PiN diodes. This implies that SEB is also not gain related, which is also confirmed in this study due to the temporary gain suppression caused by large injected charge. It was also shown that SEB is independent of pre-irradiation dose, as the same outcomes were seen for devices irradiated to different fluences. Finally, it was demonstrated that SEB is a threshold phenomenon, where more deposited charge and energy lead to lower voltage thresholds to cause SEB. While the aforementioned studies uncovered several key aspects of the SEB mechanism in sensors, this must be validated with a systematic study using heavy ion beams which are more representative of detector operational conditions.

So far, beam-induced SEB has been observed in LGADs and PiN diodes when the electric field applied to the sensor exceeds approximately 12 V/$\upmu$m, where the exact threshold depends on active thickness \cite{HGTD_TB, Torino_SEB}. This is normally not a problem for non-irradiated thin LGADs, as they experience breakdown at voltages well below this critical electrical field of 12 V/$\upmu$m. However, in high-radiation environments, radiation causes gain loss and worsens performance, and this is compensated for by increasing operational voltage. As operational voltage is increased, a larger average electric field is created in the bulk and it is more likely the sensor will experience SEB. While Refs.~\cite{HGTD_TB, Torino_SEB} studied MIP-induced SEB, studies must be extended to higher released energies for further understanding.

\begin{figure}[htbp]
    \centering
    \includegraphics[width=.75\linewidth]{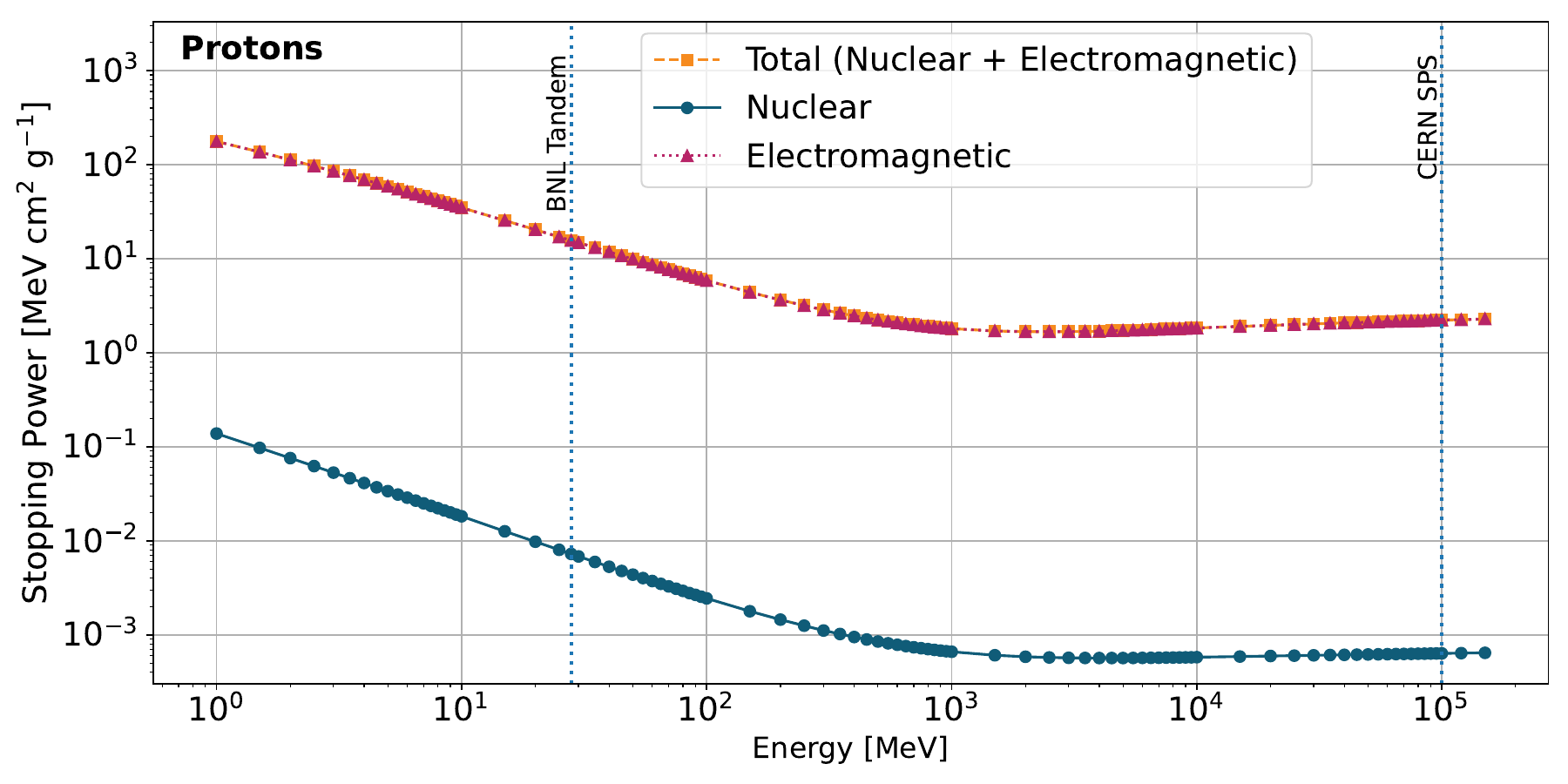}
    \caption{Comparison of electromagnetic and nuclear stopping power from SRIM simulation of protons in Silicon, computed with \cite{srniel_calculator, srniel_calculator_nuclear}. Vertical lines are included at 28 MeV protons and 100 GeV protons to show the increased stopping power expected at the Brookhaven National Laboratory Tandem van de Graaff accelerator, and CERN SPS facility.}
    \label{fig:stoppingPower}
\end{figure}

To build upon the previously described SEB studies, in this work, we report a systematic SEB study of PiN diodes and LGAD sensors. All sensors are pre-irradiated in order to reach expected SEB voltages, and then exposed to 28 MeV protons and $\mathcal{O}$(100) MeV heavy ions (carbon, oxygen, iron, and gold) produced at the Brookhaven National Laboratory Tandem van de Graaff accelerator \cite{tandem, tandem2}. By studying interactions covering a wide range of stopping power, we directly probe the response of LGADs to extreme ionization densities and investigate the interplay between stopping power, operating bias, and device failure. In particular, Fig. \ref{fig:stoppingPower} shows stopping power as a function of Energy for protons in Silicon, where we expect more stopping power at 28 MeV compared to previous studies which used $\mathcal{O}$(100) GeV protons. Similar plots for other ion species are shown in~\ref{appendix:StoppingPowerPlots}.

Sensors used for particle detection, such as LGADs and PiN diodes, operate under closely analogous conditions. When operating under high voltages, as is done after detectors are subjected to large radiation doses, they become intrinsically vulnerable to localized runaway processes. It has been observed that MIPs can induce SEB, and therefore have the possibility to generate sufficient charge density to perturb the electric field. This invites the question of whether particles with high stopping power can cause different damage phenomenology.

The results of this work show various damage and mortality modes caused by operational procedures, including the application of very high voltage with no beam, and from both MIP and non-MIP particles. This includes experimental evidence of a change in SEB-like damage phenomenology as a function of stopping power, consistent with SEB-like behavior. This observation also confirms heavy-ion-induced single-event effects as a relevant reliability concern for LGAD-based timing detectors and defines critical constraints on sensor design and operating conditions for current and future high-radiation experiments.
\section{Sensor preparation}
\label{sec:Devices_Tested}

\begin{figure}[!ht]
    \centering
    \includegraphics[width=0.35\linewidth]{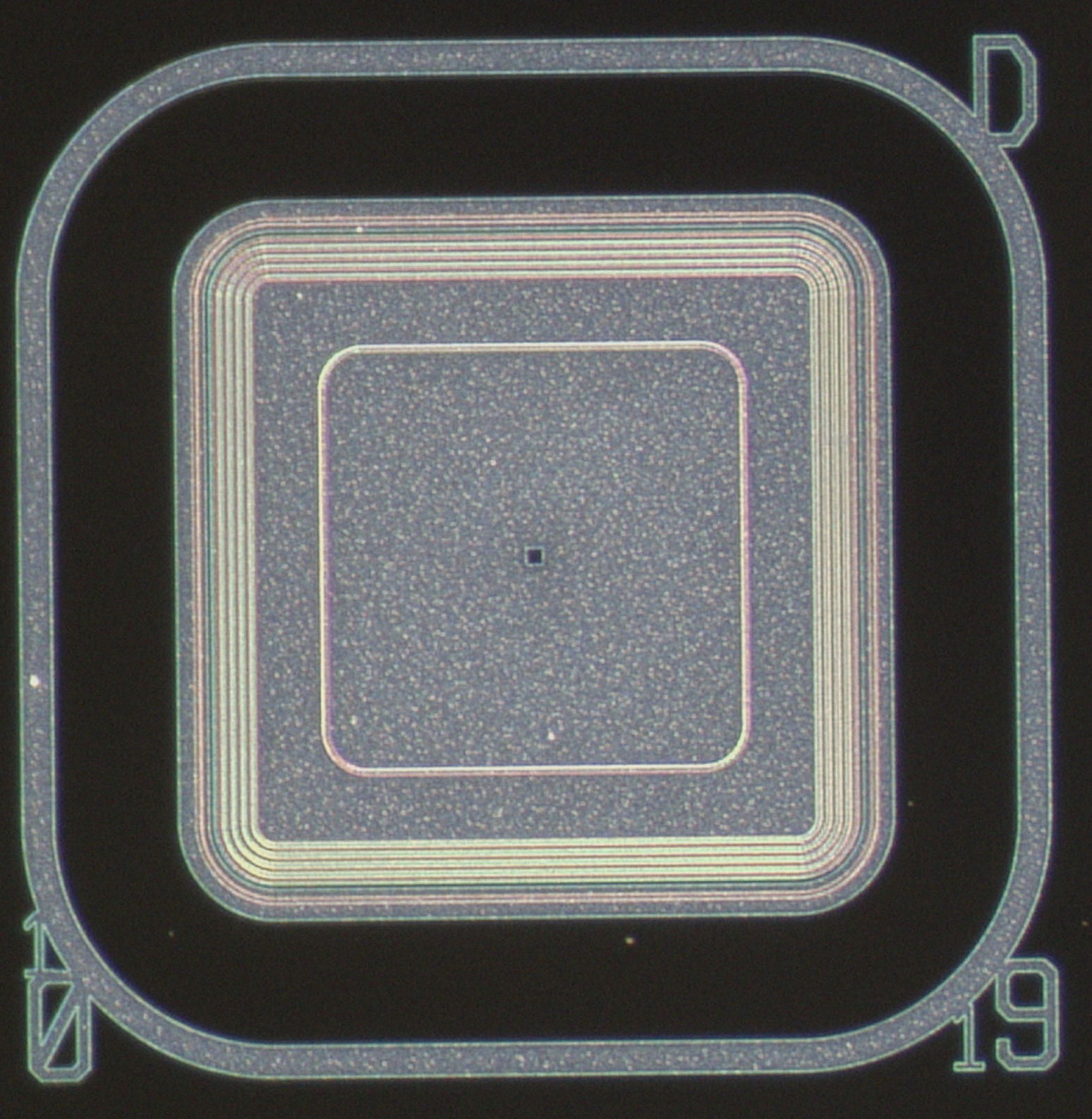}
    \includegraphics[width=0.55\linewidth]{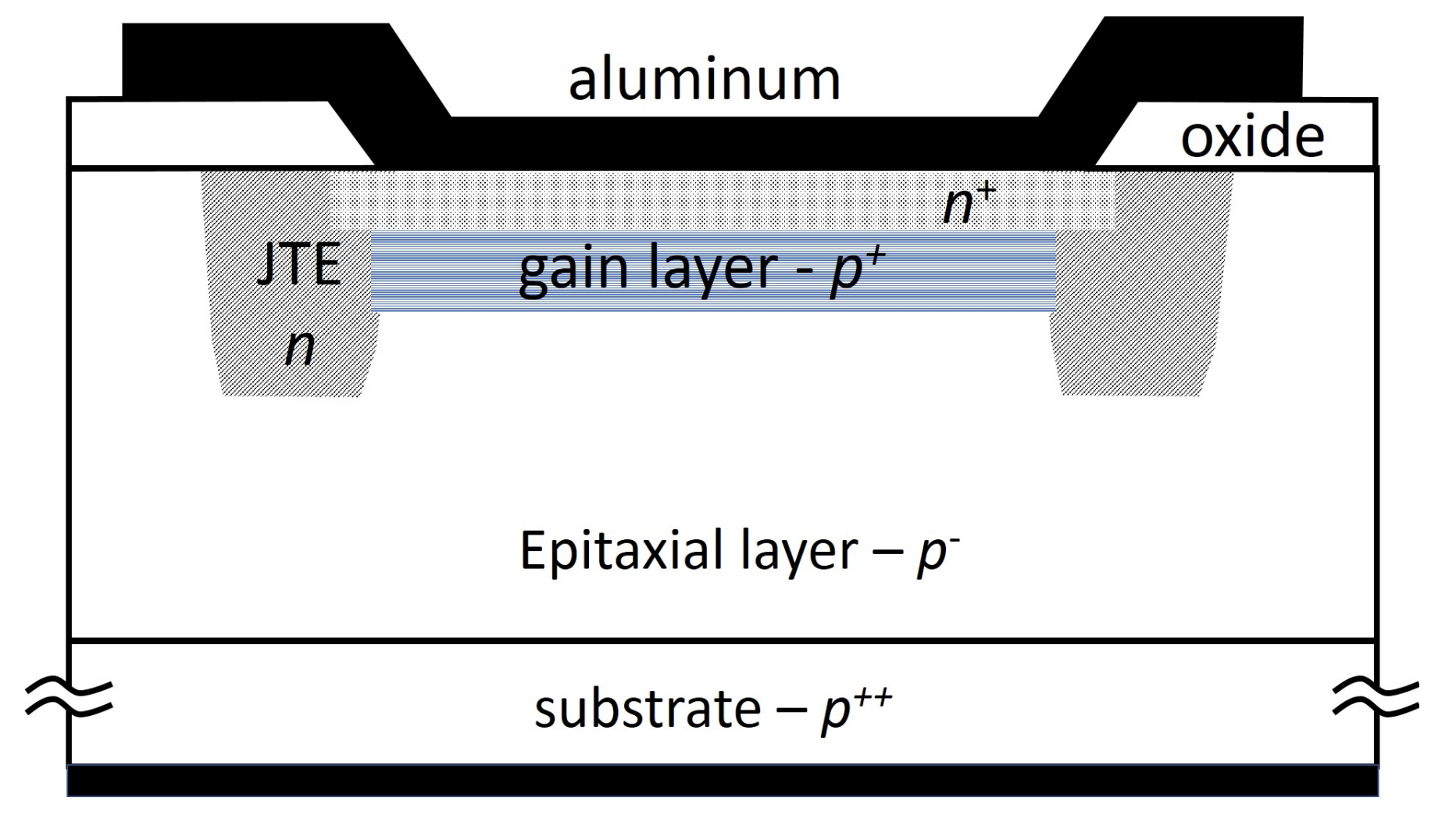}
    \caption{\textit{(left)} Picture of an example BNL-fabricated device used in this work prior to testing with beams: diode with metal covering the active area. \textit{(right)} Diagram of the structure of a BNL fabricated LGAD. The PiN diodes used in this study share an identical structure, but lack the gain layer.}
    \label{fig:LGADphoto}
\end{figure}

The devices under test (DUT) consist of multiple variants of Silicon sensors, diodes and LGADs, fabricated in the class-100 Silicon clean-room run by the Instrumentation Department at Brookhaven National Laboratory (BNL)~\cite{LGADBNL}. They consist of both LGADs and PiN diodes, with active layer thicknesses of 20, 30, and 50 $\upmu$m, all with 300 $\upmu$m thick substrates. The geometry is the same for both diodes and LGADs, the only difference being the presence of a 1.2~mm x 1.2~mm gain layer under the 1.3~mm x 1.3~mm pad, as shown in Fig.~\ref{fig:LGADphoto}
. Table \ref{tab:sensors_under_test} summarizes the sensors used in this study.

In order to emulate the behavior of these sensors after prolonged operations in high-radiation environments, they must be pre-irradiated. In particular, pre-irradiation is necessary to increase their breakdown voltages to their expected SEB thresholds, to test whether these sensors experience SEB around the expected threshold of 12 V/$\upmu$m. This was achieved at the Rhode Island Nuclear Science Center (RINSC) \cite{RINSC} by irradiating the samples up to a fluence of $1.5\cdot10^{15}\mathrm{n_{eq}/cm^2}$, using neutrons produced by a 2 MW light water cooled MTR Pool Type reactor. Each sensor was affixed to one of four faces of a wooden block, alongside iron foils used to measure the received dose, and inserted into the reactor using a pneumatic tube system. These blocks spent 80 minutes exposed to the reactor core to receive the $1.5\cdot10^{15}\mathrm{n_{eq}/cm^2}$ target fluence. The orientation of the sensors relative to the core can vary the dose received by up to $15\%$. By targeting $1.5\cdot10^{15}\mathrm{n_{eq}/cm^2}$, these results are comparable to other MIP-based studies of SEB in LGADs \cite{HGTD_TB, Torino_SEB}.

\begin{table}[htb]
    \centering
    \begin{tabular}{l|c|c|c|c} \hline
    Type    & \makecell{Active \\thickness [$\upmu$m]} & \makecell{Unirradiated \\breakdown\\voltage [V]} & \makecell{Expected SEB \\threshold [V]} & \makecell{Number of \\devices} \\\hline\hline
    PiN     & 30 & $>$ 300 & 360 & 10 \\\hline
            & 20 & 100 & 240 & 22  \\
    LGAD    & 30 & 130 & 360 & 26  \\
            & 50 & 200 & 600 & 14  \\ \hline
    \multicolumn{1}{c}{} & \multicolumn{1}{c}{} & \multicolumn{1}{c}{} &  & Total: 72 \\ \hline
    \end{tabular}%
    \caption{Number of devices under test per type and thickness.}
    \label{tab:sensors_under_test}
\end{table}

After pre-irradiation, all devices were baked at 60$\degree$C for 80 minutes to induce a controlled annealing and bring the devices to a stable state - a standard of the CERN-RD50 project \cite{Hartmann2024}. This ensures that self-annealing is not a factor and will not influence in-situ measurements and results.
\section{Experimental setup}

\begin{figure*}[tbp]
    \centering
    \includegraphics[width=\linewidth]{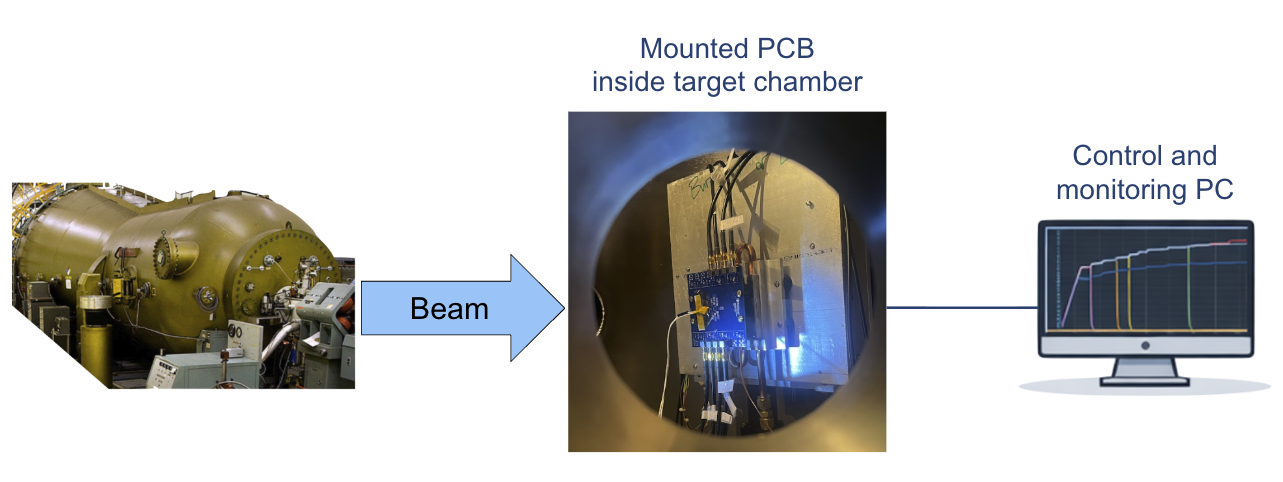}
    \caption{Test beam setup overview, from left to right: The Van de Graaff electrostatic accelerator, which provides beam to the test chamber. Pictured inside the test chamber is the mounted PCB. HV and cooling are provided to the target chamber, and signals are read out and connected to a control and monitoring PC.}
    \label{fig:setup}
\end{figure*}

Sensors are glued on a custom printed circuit board (PCB), which is mounted on a support structure. The support structure is placed inside a vacuum chamber, and sensors are biased and monitored in-situ. An overview of the test beam setup is shown in Fig. \ref{fig:setup}, and its various components are described in the following.

\subsection{Sensor bias and readout}
The DUTs are biased using a CAEN~R8033DN programmable high-voltage power supply (HV-PSU), capable of providing up to 4 kV of negative voltage to each of its 16 channels independently. Eight channels of the HV-PSU are used to independently bias eight sensors, additionally providing a reading of the current drawn by each sensor pad with a resolution of 5 nA. The other eight channels of the HV-PSU are used to read out the current drawn by the guard-ring (GR) of each sensor and do not provide any voltage.

\begin{figure}[h!]
    \centering
    \includegraphics[width=.5\linewidth]{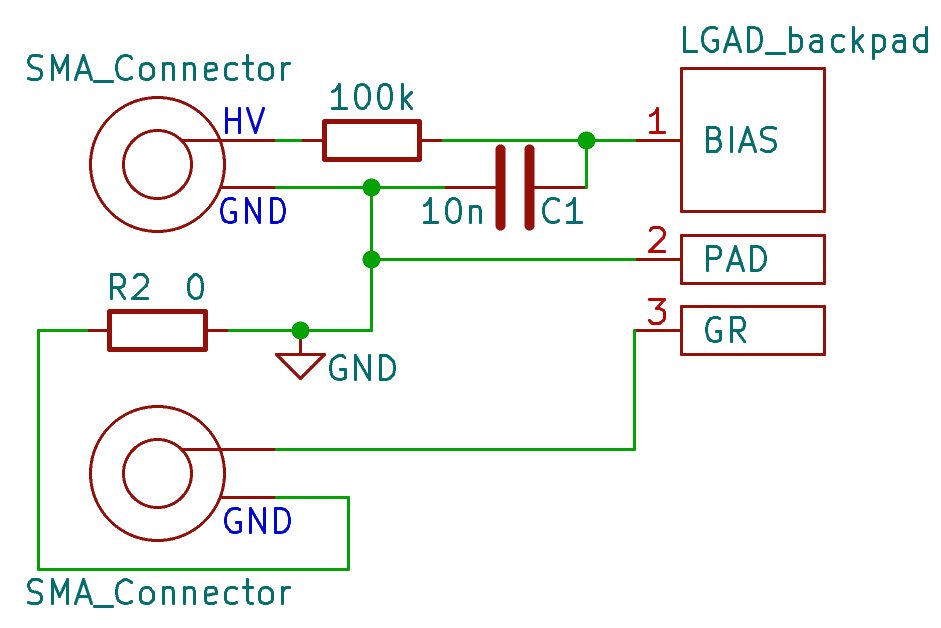}
    \includegraphics[width=.45\linewidth]{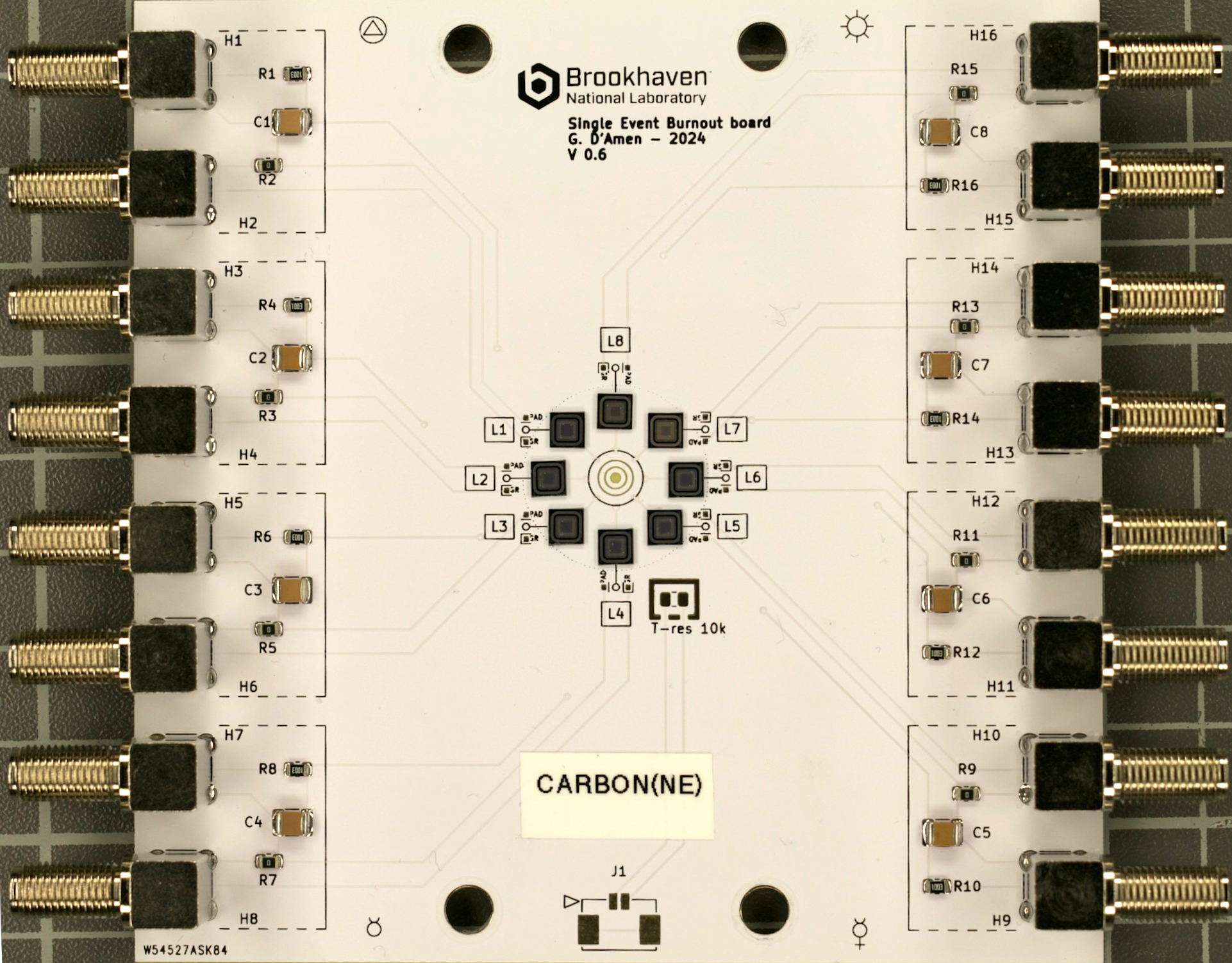}
    \caption{\textit{(left)} Schematic diagram for one of the 8 channels of the board, showing the bias line in series to each tested LGAD and diode. Each sensor is biased with negative voltage from the back (marked as BIAS in the figure). The sensors Pad (marked as LGAD) and GR are connected to ground. \textit{(right) Picture of the BNL custom-designed PCB. The 8 pads at the center are the locations where the sensors are mounted.} 
    \label{fig:board_schematic}}
\end{figure}

Each PCB hosts up to eight sensors, provides independent bias to each sensor, and is connected to the HV-PSU via 16 insulated coaxial cables. Sensors are mounted on dedicated biasing pads of the PCB using AA-DUCT 902 silver epoxy adhesive. Each sensor pad and GR are wire-bonded to dedicated bonding pads connected to independent grounds. A negative bias voltage is applied to the bottom of each LGAD sensor, while the sensor pad and GR are bonded to the ground line of independent HV-PSU channels. An RC filter composed of a 100~k$\Omega$ resistor and a 10~nF capacitor is mounted on the bias line of each sensor. A schematic diagram of the board is shown in Fig. \ref{fig:board_schematic}, together with an example picture of the full PCB.

\subsection{Target chamber}

\begin{figure*}[h!]
    \centering
    \includegraphics[width=0.8\linewidth]{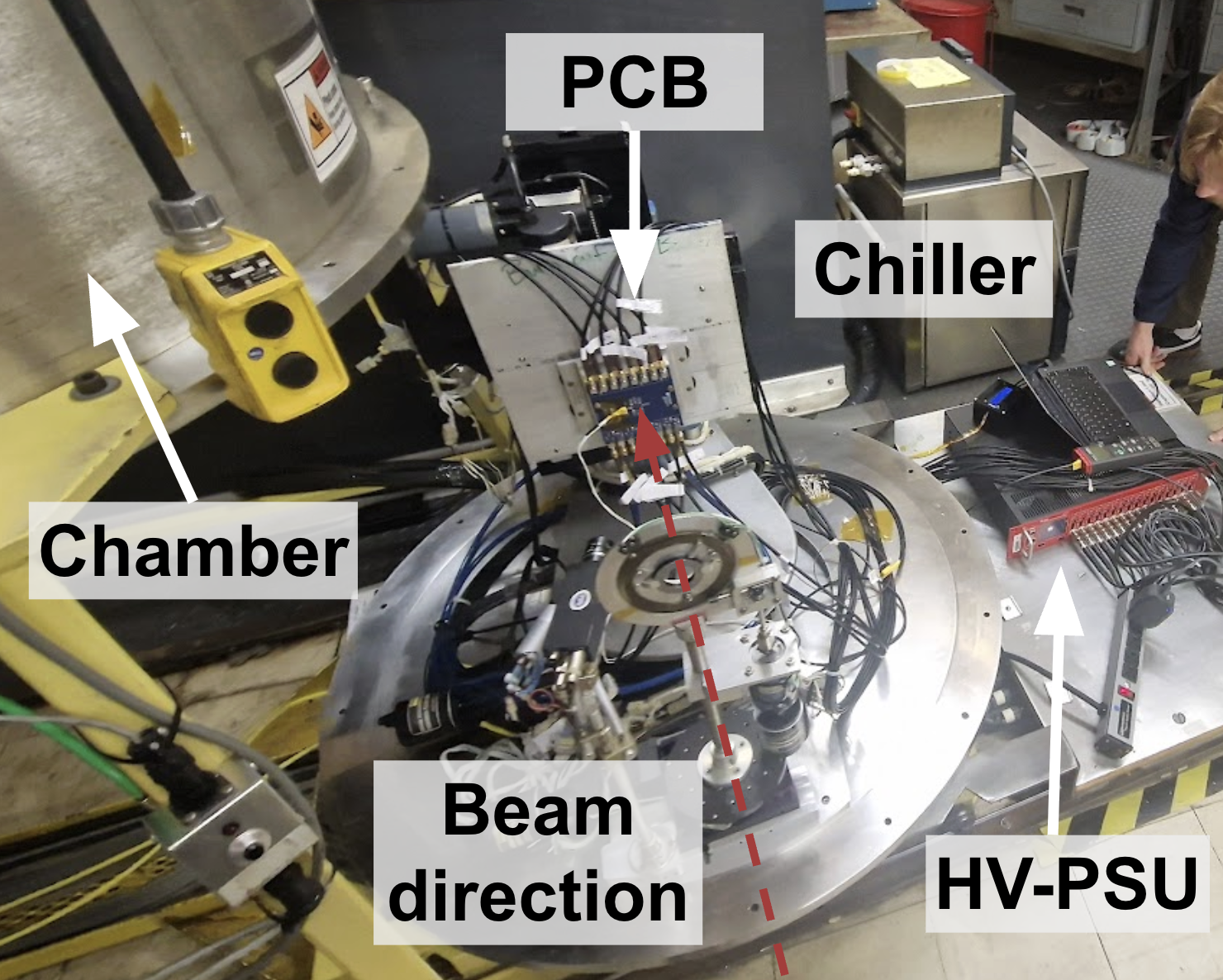}
    \caption{Image of the target setup. Pictured are the following: The test chamber, which the setup is raised into. The PCB, mounted on a metallic block for thermal control. The beam direction, where a laser alignment system is used to align the beam with the center of the PCB. The chiller used for thermal control. The HV-PSU used for sensor biasing and readout.}
    \label{fig:setup_diagram}
\end{figure*}

The sensor-loaded and HV-PSU connected PCB is mounted onto a support structure, which sits on a movable stage. This stage can be lowered, slid underneath, and then lifted up into the Tandem van de Graaff target chamber. An image of the setup is shown in Fig.~\ref{fig:setup_diagram}. 

After the stage is raised into the target chamber, vacuum is achieved by means of a pumping system to around $10^{-4}$ torr to avoid unwanted interactions between the beam and air particles, described further in ref. \cite{tandemSEU}. Heat generated by the electronics is dissipated by flowing a combination of Glycol liquid coolant and water through a chiller, through a set of cooling tubes, and then through copper pipes embedded in a metallic backplane in contact with the PCB. A thermal sensor is glued to each PCB using thermal tape in order to monitor temperature of the PCB, as a proxy for sensor temperature. The target temperature was -20$\degree$C, but due to technical challenges, board temperatures were typically around 0 $\degree$C during operations.

\subsection{Beam operations}

Once vacuum is achieved inside the test chamber, the testing PCB is aligned with the expected location of the beam, and data-taking for a given beam species can commence. In total, 6 beam types were used. They were chosen with increasing stopping power in order to study SEB as a function of released energy in the sensors, where stopping power is computed using \cite{srniel_calculator} and the parameters of the beam. These beam types are summarized in Tab.~\ref{tab:beam_species}.

\begin{table}[htbp]
\centering
    \begin{tabular}{c|c|c|c|c|c}
    Beam species    &  \makecell{Energy\\ {[MeV]}} & \makecell{Atomic Number\\(Z)} & \makecell{Mass Number\\(A)} & \makecell{Total Stopping Power\\ {[MeV cm$^2$/g]}} & \makecell{Target Fluence\\ {[ions/cm$^2$]}} \\ \hline \hline
    Proton          & 28     & 1  &  1           & 1.56 & 10$^9$  \\ 
    Carbon          & 83.5   & 6  &  12           & 1680 & 10$^7$  \\ 
    Oxygen          & 89.9   & 8  &   16          & 3330 & 10$^7$ \\ 
    Iron            & 121    & 26  &  56           & 29300 & 10$^6$ \\ 
    Gold            & 216    &  79 &  197           & 73100 & 10$^6$ \\ 
    Gold            & 332    &  79 &   197          & 84300 & 1-5$\times$10$^6$   \\ 
    \end{tabular}

\caption{Beam species, their associated energies, stopping powers, and target fluences. Stopping power is computed using \cite{srniel_calculator} for a given beam species and energy.\label{tab:beam_species}}
\end{table}

The beam operations procedure is as follows: 

\begin{itemize}
    \item[1)] With beam off, voltage was ramped up for all sensors in steps of 5-25V depending on sensor thickness and proximity to expected SEB voltage. When operating boards with multiple sensor thicknesses, voltages are increased independently.
    \item[2)] While at constant voltage, the beam was injected into sensors until target fluence (ranging from $10^{6}$-$10^{9}$ ions/cm$^2$ depending on beam species) was achieved, as shown in Tab.~\ref{tab:beam_species}.
\end{itemize}

During step 2, if a sensor experiences high current which goes above a set threshold of the HV-PSU, this channel's voltage is automatically ramped down by the HV-PSU. In cases where the current suddenly spikes while beam is on, this sensor is marked as a candidate for offline investigation. After step 2, this procedure repeats until all sensors were damaged or reached breakdown.
\section{Results}
\label{sec:results}

This section describes the post-beam analysis, categorization, and results of the 72 tested sensors. Out of the 72 sensors, 44 sensors reached their channel's threshold current without a spike in current or a signature that may be consistent with SEB, limiting the maximum voltage they could be tested at. These 44 sensors did not experience permanent damage or SEB.

\subsection{Post-beam analysis}

\begin{figure*}[htbp]
    \centering
    \includegraphics[width=\linewidth]{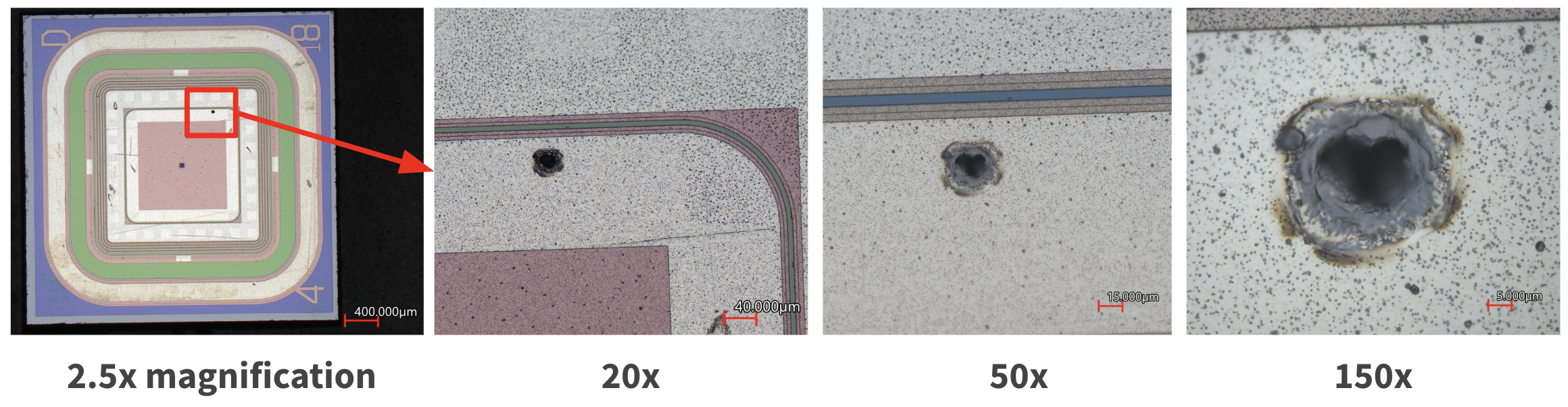}
    \caption{Example of a visual inspection of an SEB candidate sensor with a crater in the active region at 2.5x, 20x, 50x, and 150x magnifications, from left to right, respectively.}
    \label{fig:visual_inspection_example}
\end{figure*}

For each sensor candidate, i.e. those which showed a spike in current, or steady increase in leakage current at constant voltage, additional measurements and inspections were performed. First, visual inspection was performed with metrology measurements taken using a Keyence VK-X 1100 3D laser scanning confocal microscope. In 11 cases, this inspection identified craters formed on the surface of the sensor. An example of the visual inspection performed with the microscope is shown in Fig.~\ref{fig:visual_inspection_example}, and its associated metrology measurement at 150x magnification is shown in Fig.~\ref{fig:metrology_example}. Further details on metrology results can be found in \ref{appendix:MetrologyResults}.

\begin{figure}[htbp]
    \centering
    \includegraphics[width=.6\linewidth]{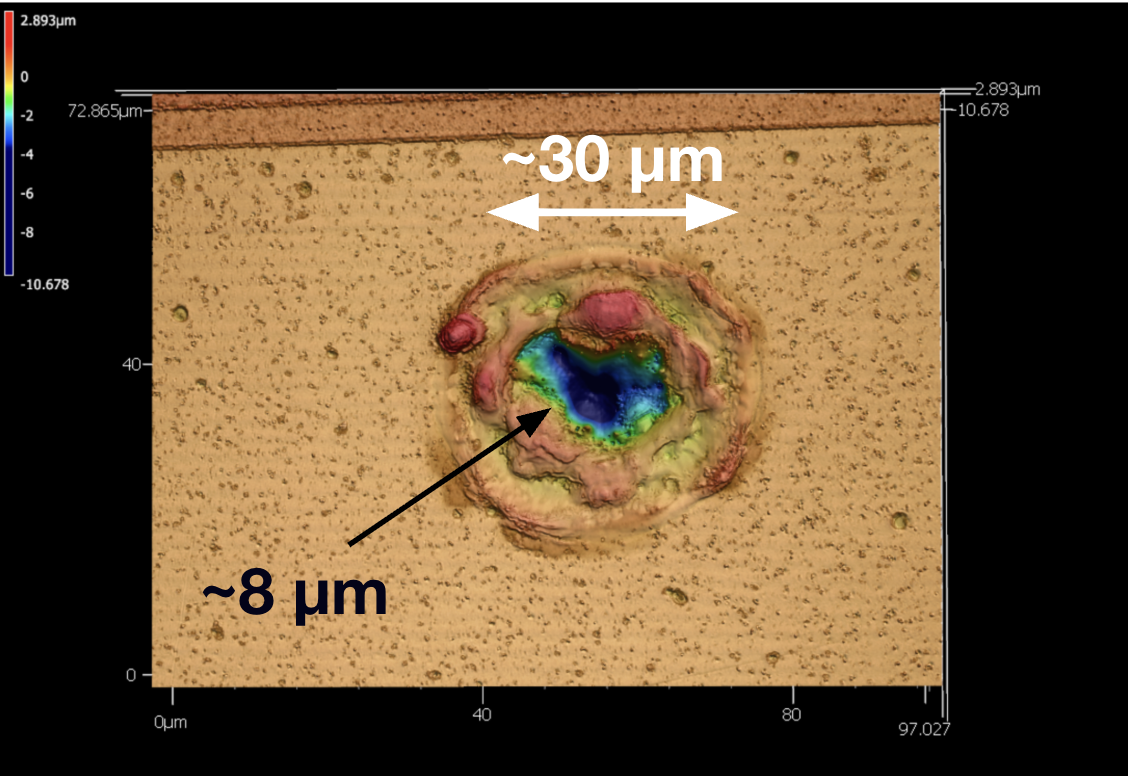}
    \caption{Example of a metrology measurement of an SEB candidate sensor around the crater area, the same as that shown in Fig.~\ref{fig:visual_inspection_example}.}
    \label{fig:metrology_example}
\end{figure}

Post-beam I-V measurements, in which sensor leakage current is measured as a function of bias voltage, were also taken for SEB candidate sensors in order to determine if the sensor was still functional. These measurements, along with visual inspection and metrology measurements, as well as comparisons to in-situ I-V information were taken into account when categorizing the sensors as described in the following.

\subsection{Phenomenology and categorization}

Several types of damage phenomenology are observed in the in-situ voltage and current measurements, in metrology, and in post-beam I-V measurements. During beam operations, sensors show spikes in current with either beam on or beam off. In both cases, upon re-biasing the sensors, the current spikes very high at low voltages, indicating damage. Additionally, some sensors show a steady increase in current as beam is injected. In this case, re-biasing yields a higher current than before beam was injected, indicating damage. Metrology measurements showed two types of craters: Craters that are circular in size, usually with widths around 30 $\upmu$m, and depths around 8 $\upmu$m. Several cases of craters near the guard ring are observed, with non-circular shapes. I-Vs were taken of candidate sensors after the test beam campaign to confirm whether sensors have differing leakage current compared to in-situ current measurements.

Combining these measurements, sensors fall into three distinct damage categories:

\begin{itemize}
    \item[]Category 1: SEB candidate.
    \item[]Category 2: Damage from high current, no beam.
    \item[]Category 3: Damage from beam and/or electrical effects.
\end{itemize}

\begin{figure*}[tbp]
    \centering
    \includegraphics[width=0.65\linewidth]{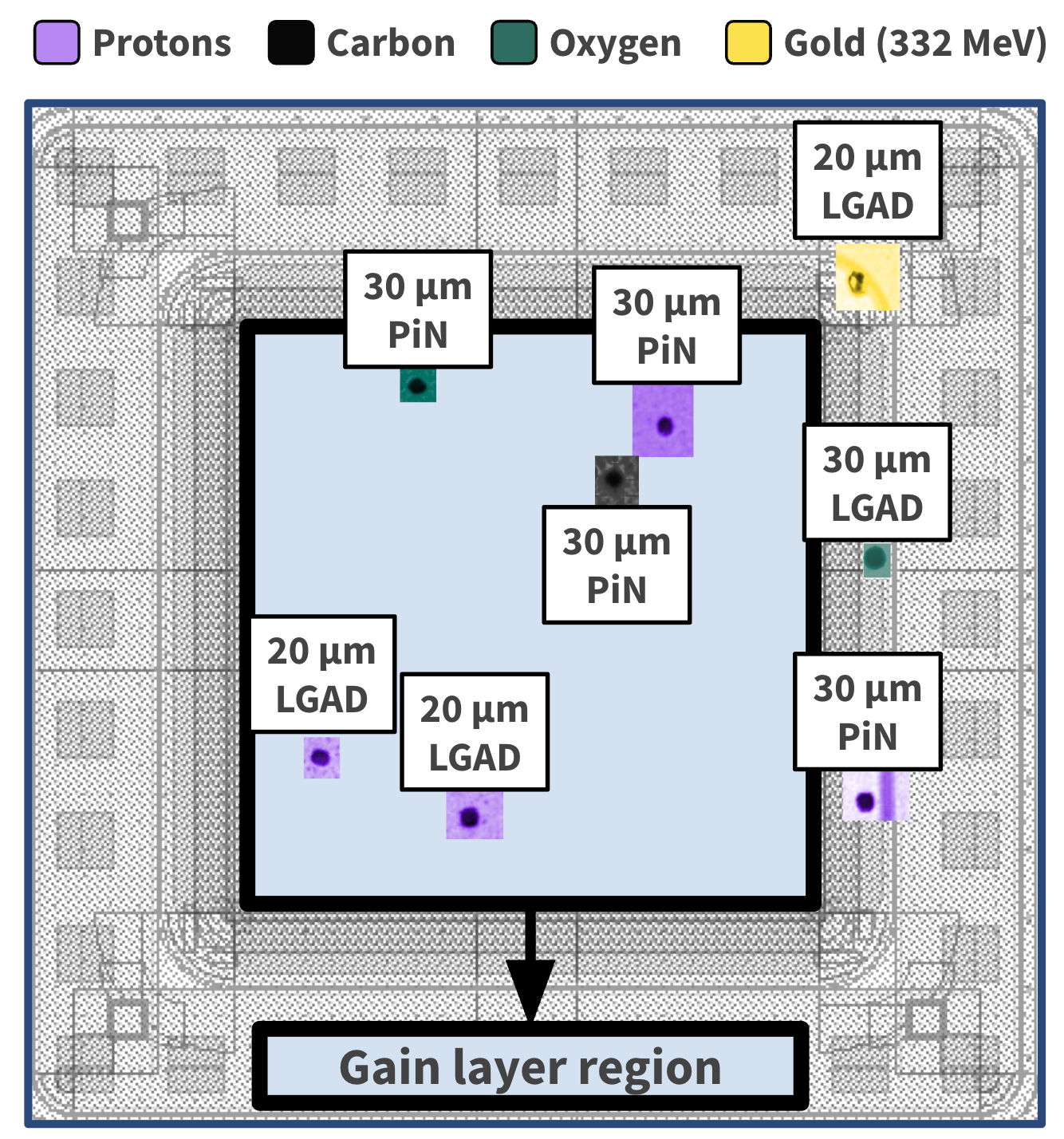}
    \caption{Location of craters on Category 1 sensors (SEB candidates). The gain layer region for LGADs is shaded in blue.}
    \label{fig:cat_1_craters}
\end{figure*}

For Category 1 sensors: a spike in current is observed while beam is on at a fixed voltage, and, upon re-biasing the sensor, the current spikes very high at low voltages, indicating sensor damage. The locations on the sensor surface of the Category 1 craters are shown in Fig.~\ref{fig:cat_1_craters}. Within the statistical uncertainty of the dataset used, there is no clear spatial dependence of craters. Craters are seen for both protons and heavy ions, and there is no clear dependence of crater position on sensor type and thickness. For most observed Category 1 craters, similar dimensions were measured: circular shapes with $\approx$ 30 $\upmu$m diameter, and depth of about 8 $\upmu$m.

\begin{figure}[tbp]
    \centering
    \includegraphics[width=.8\linewidth]{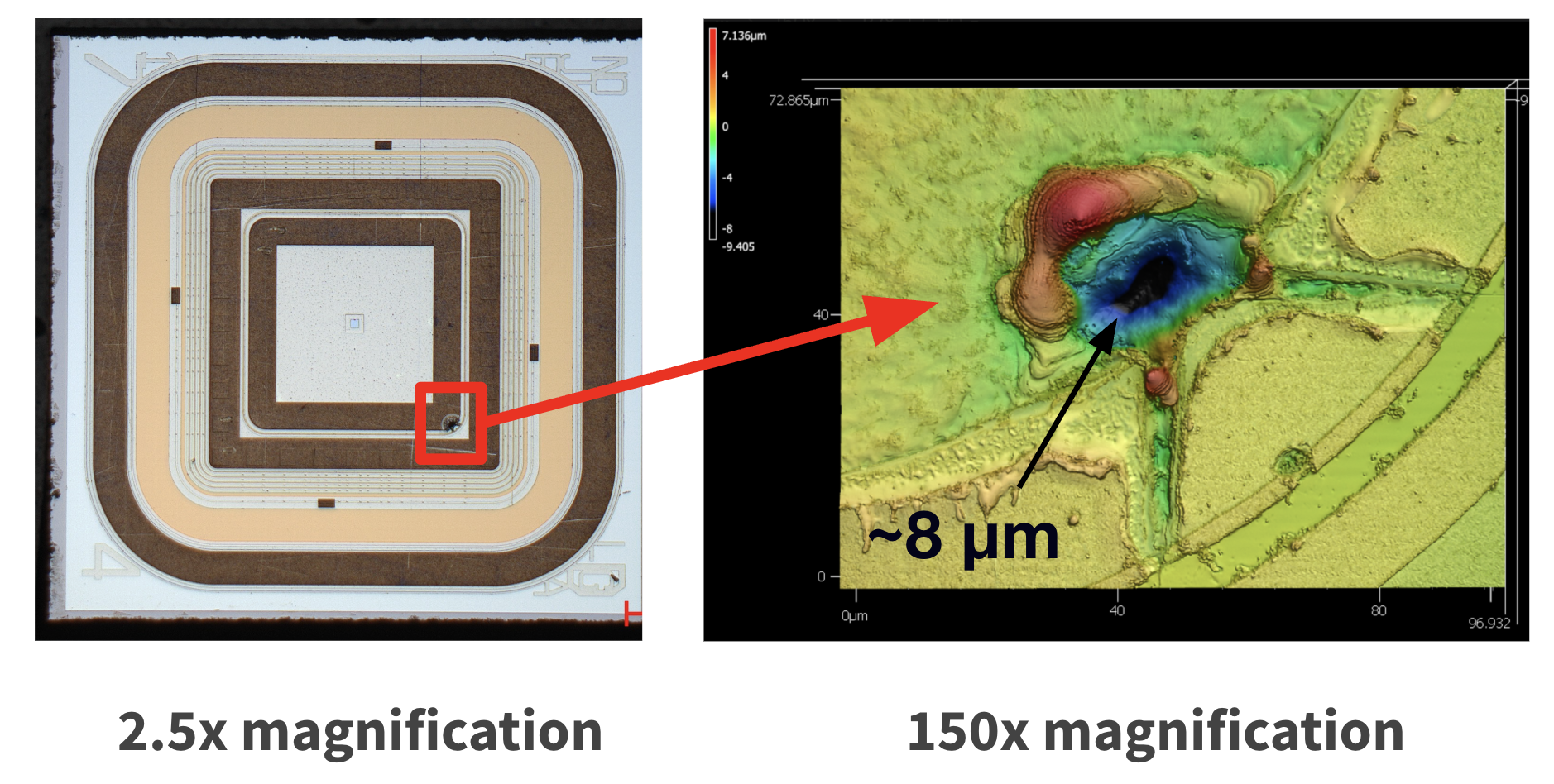}
    \caption{Example of a category 2 crater.}
    \label{fig:cat_2_crater}
\end{figure}

Category 2 sensors show a relatively low current until a breakdown-like curve is observed when reaching a high voltage with no beam. Upon re-biasing, these sensors typically show very high current before reaching high voltages, indicating that damage was caused even without beam, but likely by exposing the sensor to very high current. For Category 2 sensors, 3 sensors show craters, all of which are near the GR area. An example is shown in Fig.~\ref{fig:cat_2_crater}. Category 2 craters have very similar depth to Category 1 craters.

\begin{figure}[htbp]
    \centering
    \includegraphics[width=.75\linewidth]{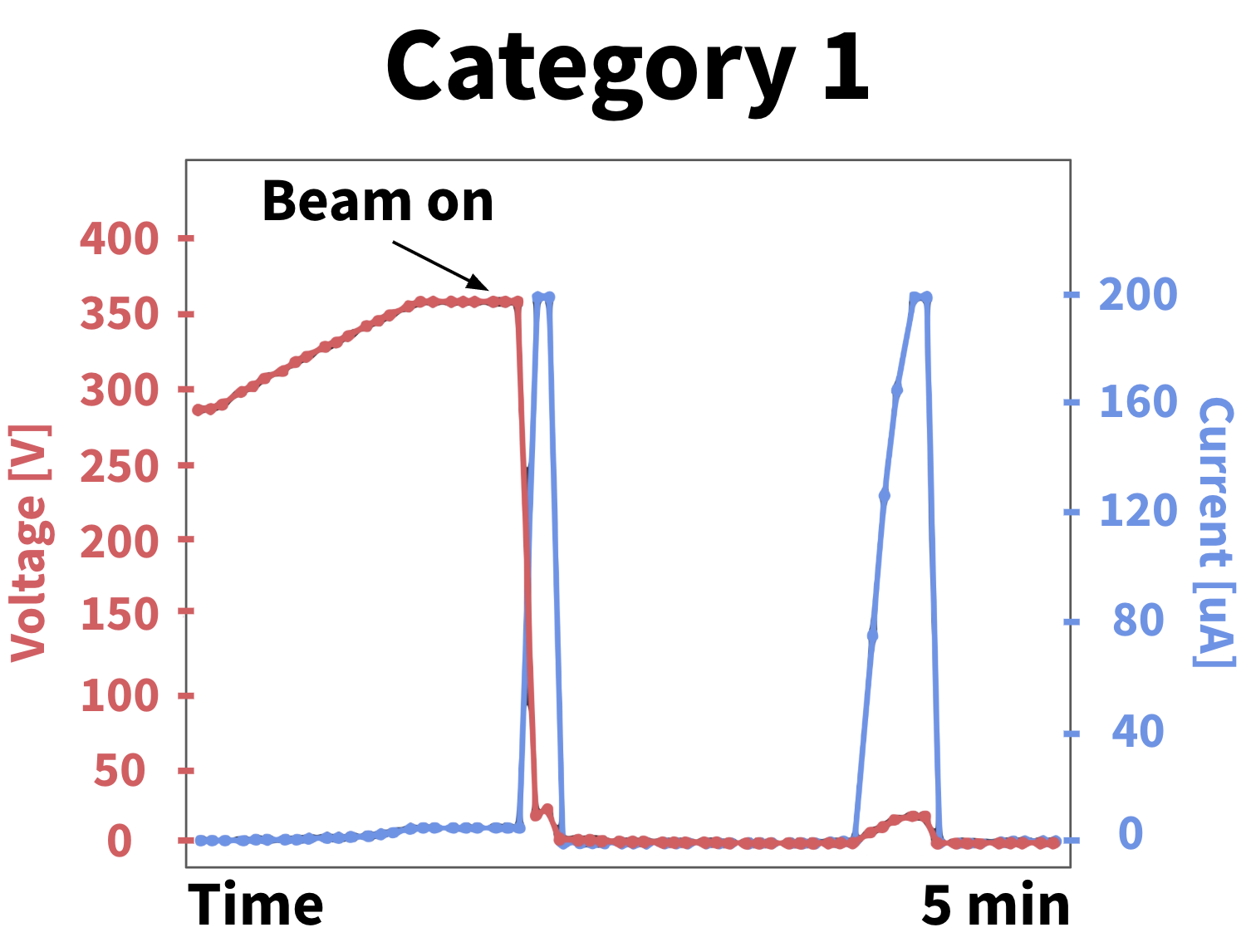}
    \caption{In-situ voltage and current measurements typical of Category 1 sensors (``SEB candidates"), showing a spike in current while the beam is on and sensors are kept at fixed voltage.}
    \label{fig:cat_1_insitu}
\end{figure}

\begin{figure}[htbp]
    \centering
    \includegraphics[width=.75\linewidth]{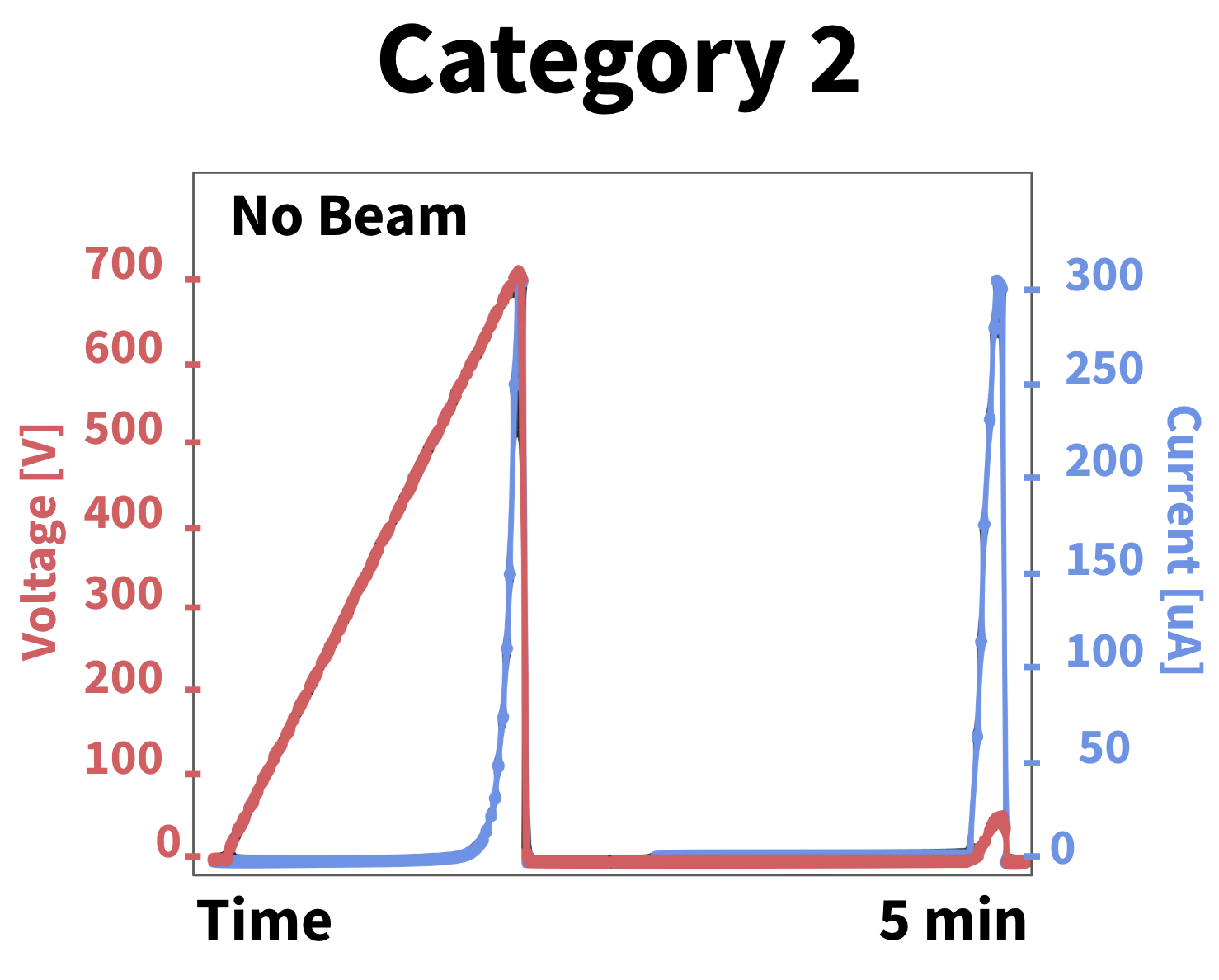}
    \caption{In-situ voltage and current measurements typical of Category 2 sensors (``Damage from high current, no beam"), showing rapidly increasing current while ramping up the voltage with the beam off. This figure shows an example for a 50 $\mu$m LGAD.}
    \label{fig:cat_2_insitu}
\end{figure}

\begin{figure}[htbp]
    \centering
    \includegraphics[width=.75\linewidth]{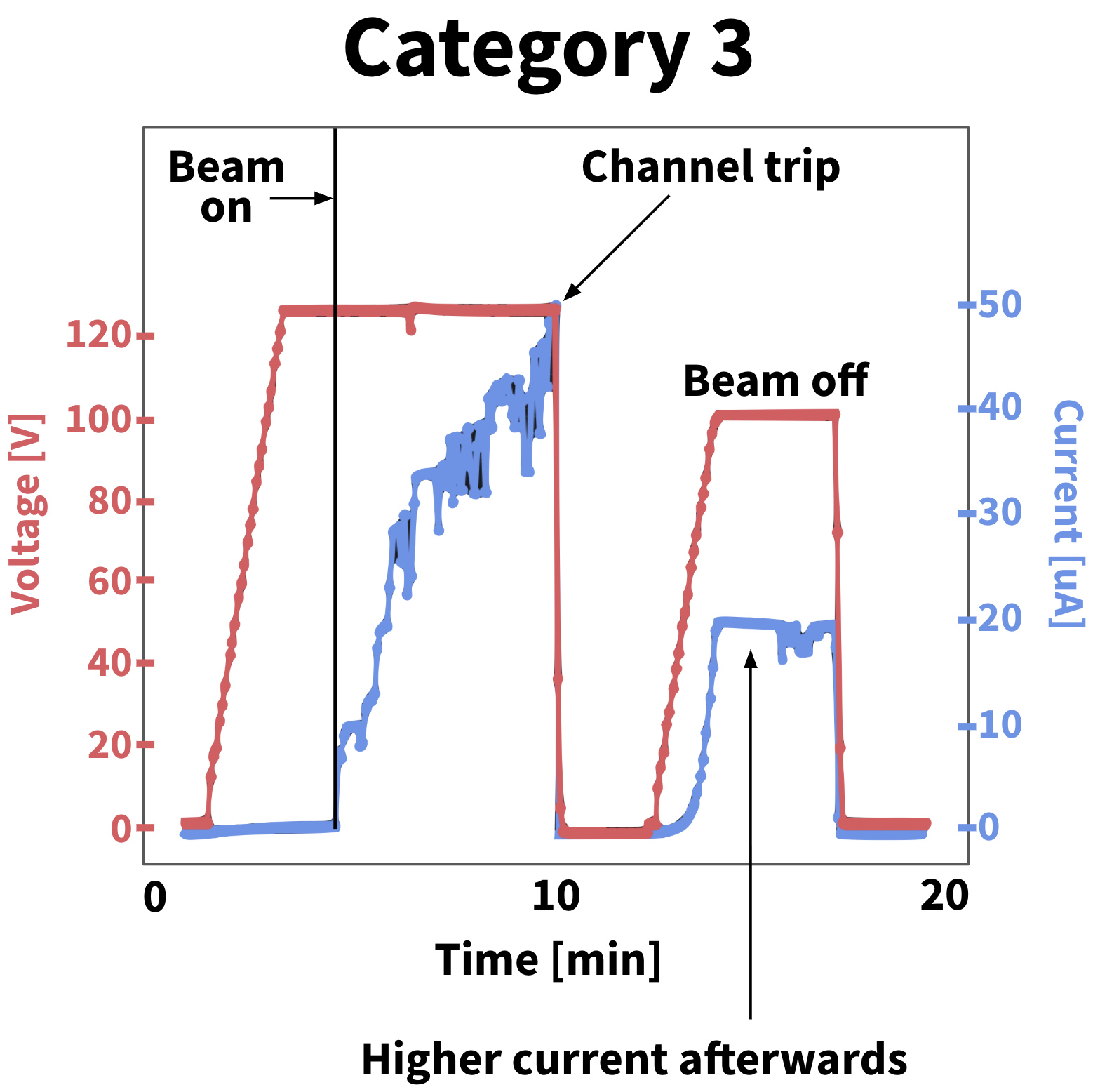}
    \caption{In-situ voltage and current measurements typical of Category 3 sensors (``Damage from beam and/or electrical effects"), showing a steady increase in leakage current as soon as the beam is injected.}
    \label{fig:cat_3_insitu}
\end{figure}

Category 3 shows a steady increase in leakage current as soon as the beam is injected, leading to the channel tripping when a set maximum current is reached. Upon re-biasing, a higher current is observed at a lower voltage, potentially indicating damage to the crystalline lattice of the devices under test. The in-situ voltage and current measurements which were typical of each category are shown in Figs.~\ref{fig:cat_1_insitu}, \ref{fig:cat_2_insitu}, and \ref{fig:cat_3_insitu} respectively.


\begin{figure*}[tbp]
    \centering
    \includegraphics[width=\linewidth]{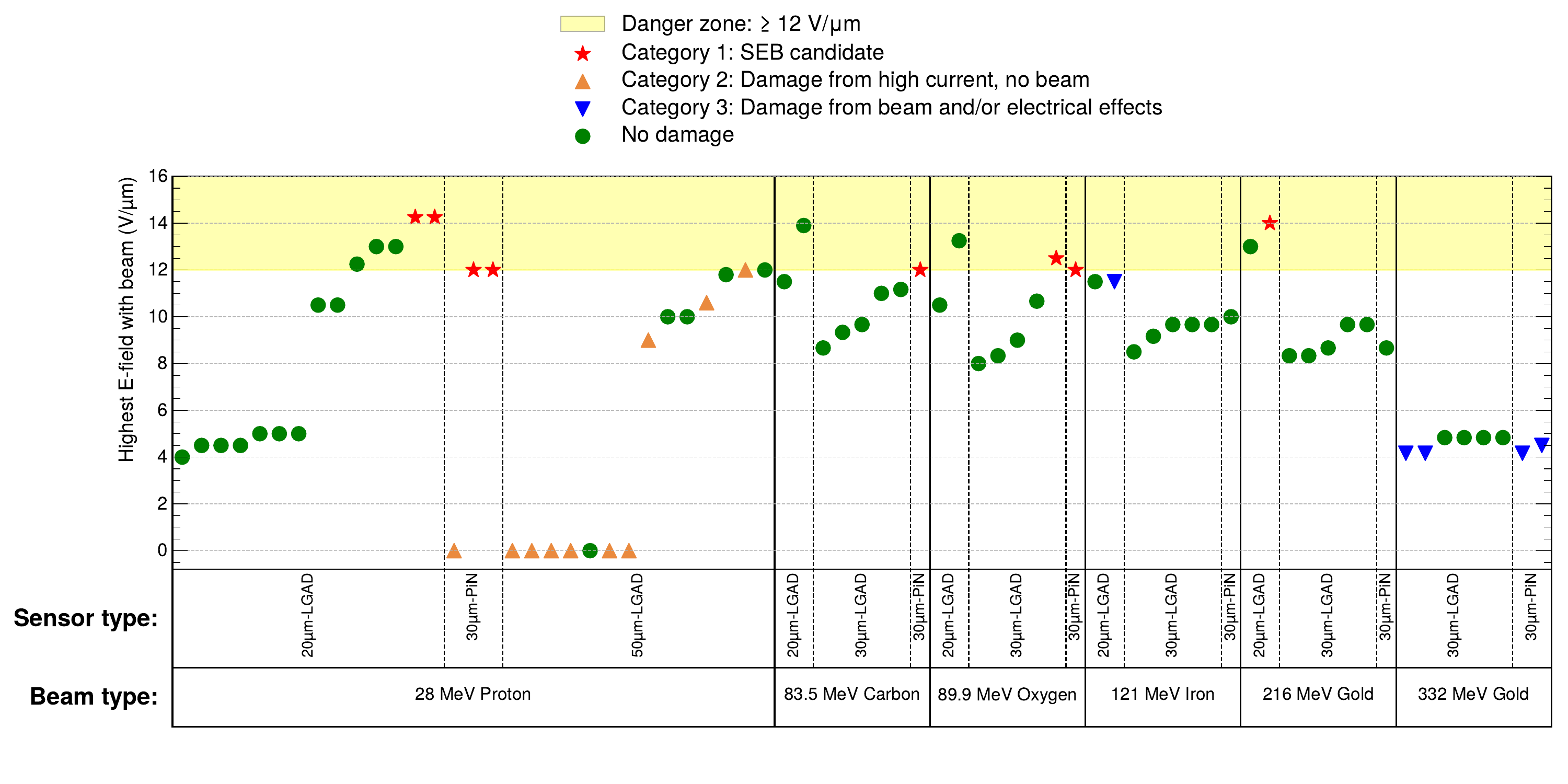}
    \caption{Results summary. Each data point represents the last bias voltage before death or breakdown of one sensor, displaying the highest electric field that was reached for a given sensor with beam on, and the category the sensor falls into. Sensors are separated by type (LGAD/PiN, active thickness), and different types tested for each beam type are shown.}
    \label{fig:all_results_plot}
\end{figure*}

\begin{sidewaysfigure}
    \centering
    \vspace{8cm}
    \includegraphics[width=\linewidth]{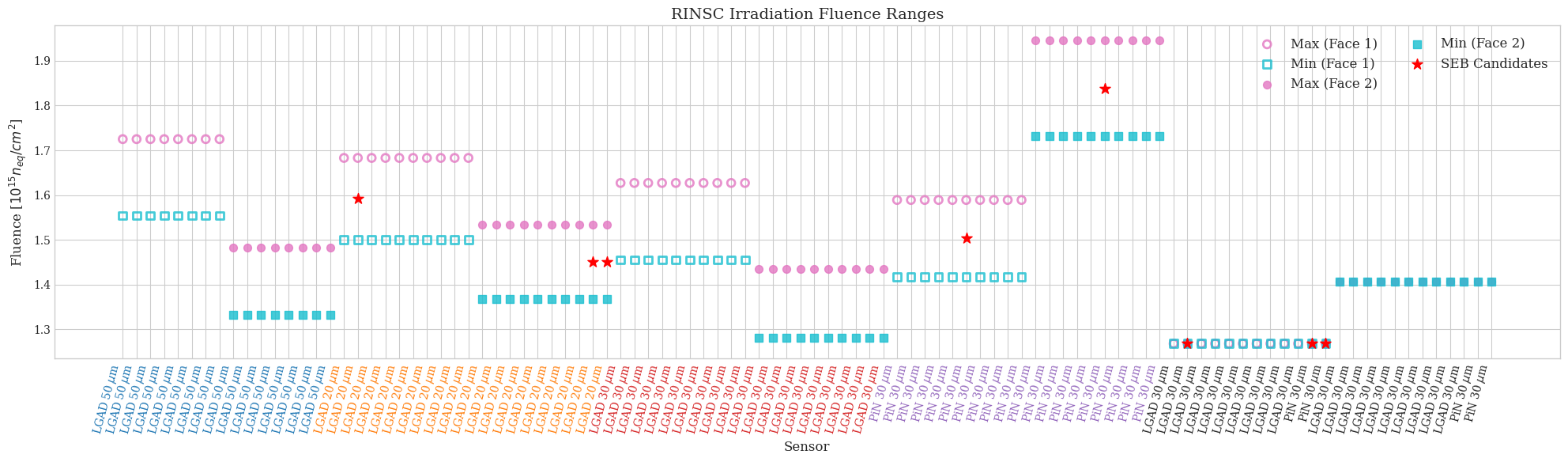}
    \caption{A summary of the RINSC pre-irradiation dose received by all sensors. For each sensor, a dose range is shown based on the minimum and maximum estimate doses received by the block the sensor was affixed to, as described in Sec.~\ref{sec:Devices_Tested}. SEB candidate sensors falling into Category 1 are highlighted with a red star. Faces 1 and 2 correspond to the two sides of the rabbit sticks used at the RINSC facility.}
    \label{fig:RINSC_dose_plot}
\end{sidewaysfigure}

Out of the 72 total sensors, 23 sensors fell into one of these three categories, while the remaining sensors did not exhibit any behavior consistent with sensor damage. A summary table with all candidate sensors falling into these categories is shown in Tab.~\ref{tab:categorization_table}, the total set of results is shown in Fig.~\ref{fig:all_results_plot}, and a summary of the RINSC pre-irradiation doses with Category 1 sensors marked is shown in Fig.~\ref{fig:RINSC_dose_plot}.

\setlength{\arrayrulewidth}{1.2pt}

\begin{table*}[htbp]
\centering

\resizebox{\textwidth}{!}{%
    \begin{tabular}{c|c|c|c|c|c}
    Category & Beam type      & Sensor type  & Sensor thickness [$\mu$m] & \makecell{Estimated pre-irradiation\\fluence [$10^{15}\mathrm{n_{eq}/cm^2}$]} & Damage threshold [V/$\mu$m] \\ \hline \hline

     & Protons               & PiN      & 30 & 1.56 & 12    \\
     & Protons                & PiN      & 30 & 1.91 & 12    \\
     & Protons                & LGAD      & 20 & 1.37 & 14.25   \\
    1: SEB candidate & Protons                & LGAD      & 20 & 1.37 & 14.25     \\
     & Carbon                & PiN      & 30 & 1.27 & 12    \\
     & Oxygen                & PiN      & 30 & 1.27 & 12     \\
     & Oxygen                & LGAD      & 30 & 1.27 & 12.5    \\
     & Gold (216 MeV)        & LGAD      & 20 & 1.5 & 14.5    \\ \hline

     & Protons               & PiN       & 30  & 1.56 & 12.5   \\
     & Protons               & LGAD      & 50  & 1.33 & 14     \\
     & Protons               & LGAD      & 50  & 1.55 & 13.2  \\
     & Protons               & LGAD      & 50  & 1.33 & 15   \\
     & Protons               & LGAD      & 50  & 1.33 & 13.4    \\
    2: Damage from high current, no beam& Protons               & LGAD      & 50  & 1.33 &  12.5   \\
     & Protons               & LGAD      & 50  & 1.55 &  15.6    \\
     & Protons               & LGAD      & 50  & 1.33 &  14.6   \\
     & Protons               & LGAD      & 50  & 1.33 & 9.2  \\
     & Protons               & LGAD      & 50  & 1.33 &  10.4   \\ \hline
    
      & Gold (216 MeV)             & PiN      & 30 & 1.91 & N/A      \\
      & Gold (216 MeV)              & PiN      & 30  & 1.91 &  N/A    \\
      & Gold (216 MeV)              & LGAD      & 30  & 1.35 & N/A    \\
    3: Damage from beam and/or electrical effects  & Gold (216 MeV)               & LGAD      & 30  & 1.54 & N/A     \\
      
      & Iron                & LGAD      & 20  & 1.37 & N/A      \\

    \end{tabular}
}

\caption{Categorization of all sensors. Each row represents one sensor, for which its final category is displayed, the beam type injected into it, sensor type and active thickness, and observed SEB threshold where applicable. \label{tab:categorization_table}}

\end{table*}

\clearpage
\clearpage


\subsection{Discussion: Confirmation of existing results}

The previously described results lead to several key findings. For Category 1 sensors, the approximate 12 V/$\upmu$m threshold is confirmed. Furthermore, the results appear consistent with those in Ref. \cite{Torino_SEB} for 20 $\upmu$m thick sensors, which show higher SEB thresholds for thinner sensors. Thresholds of approximately 14.25-14.5V/$\upmu$m are observed for 20 $\upmu$m thick sensors, while a lower threshold of $\approx$12-12.5 V/$\upmu$m is observed for 30-50 $\upmu$m sensors. This is also consistent with Refs.~\cite{Laser_Jinst, LASTOVICKAMEDIN2022167321}, as a thickness dependent threshold is observed.

Fig~\ref{fig:RINSC_dose_plot} shows a spread in pre-irradiation dose for Category 1 sensors, indicating that SEB likelihood does not strongly depend on the precise value of pre-irradiation dose in this range. It is also noted that both LGADs and PiN diodes were affected by all three mortality categories including SEB, suggesting that these mechanisms do not rely on the presence of a gain layer. Both of these observations are consistent with Refs.~\cite{Laser_Jinst, LASTOVICKAMEDIN2022167321, Torino_SEB}, which all observed that SEB does not depend on pre-irradiation dose, or the presence of a gain layer.

Within the statistics of the dataset, the locations of the Category 1 craters --- suspected to be created by SEB-like mechanisms --- do not show a strong dependence on position as a function of ion species, sensor type, and thickness. For the highest LET value from 332 MeV gold, a crater is observed near the guard ring area. More statistics are required to study if there is a dependence of crater location on LET value. The category 2 craters, produced from high current without beam, all appear near the guard ring area.

\subsection{Discussion: New results}

In addition to the above, some results of this work extend the current knowledge about mortality of LGADs and PiN diodes in operational conditions, and this work includes the first study of silicon detector mortality with oxygen, iron, and gold ion beams. By exploring a range of LET values and bias voltage values with and without beam, the distinction of three mortality modes is made in this work. In particular, it is shown for the first time that craters formed by high operational voltage without beam have very similar depth to craters caused by SEB-like mechanisms, both around 8 $\upmu$m. This implies a similarity in these damage mechanisms. Additionally, most category 1 craters are circular with $\approx$ 30 $\upmu$m diameter. For oxygen, one crater is observed to have a more oval-like shape rather than a circle, suggesting LET can potentially impact crater dimensions.

Compared to existing results, the shapes of the observed craters are sometimes similar, and sometimes different. While star-shaped craters were observed on the surface of LGADs when irradiated with protons at FNAL \cite{Hidalgo2019CMSRUN2b}, craters produced by protons in this work are circular. Given that the LET of the protons at FNAL and BNL should be very similar, this difference is currently theorized to be due to different capacitance used for the measurement circuit, and would require further dedicated study. The craters near the guard ring, one produced by 332 MeV gold, and several category 2 craters produced without beam, are similar in shape to those observed in Refs.~\cite{Laser_Jinst, LASTOVICKAMEDIN2022167321}. 

As shown in Fig.~\ref{fig:all_results_plot}, it is observed that mortality threshold is dependent on ion species, most strongly seen from the Category 3 sensors irradiated by iron and 332 MeV gold. This demonstrates there is a link between incident particle LET and damaging voltage threshold, and suggests SEB-like behavior is more likely from high LET ions, and requires a lower electric field in the bulk to trigger such behavior. This is expected based on previous femtosecond-laser-based results \cite{Laser_Jinst, LASTOVICKAMEDIN2022167321}, and previous heavy ion studies on Silicon-based power devices \cite{1589182}, but is demonstrated for the first time in Silicon LGADs.

Finally, these results provide new insights into existing simulation efforts: \cite{LASTOVICKAMEDIN2022167321} reports that 120 GeV protons deposit an energy of up to 140 MeV in the silicon bulk according to the \textsc{Geant4} simulation presented in Ref.~\cite{Heller2021LGAD}, which suggests that an energy deposit of at least 40-50 MeV is required to trigger SEB in LGAD devices. This threshold was obtained by combining the simulation of energy deposited in silicon by 120~GeV protons with the experimentally observed mortality in such conditions for irradiated LGAD devices. However, our experimental results show that 28~MeV protons can still trigger SEB while being below the threshold presented by these simulations. This work lowers that lower bound to 28~MeV, but preliminary \textsc{Geant4} simulations suggest that the energy threshold might be even lower.
\section{Conclusion}
\label{sec:conclusion}

A study to further understand SEB in Silicon sensors has been performed using the BNL Tandem van de Graaff facility. Sensors of both LGAD and PiN diode type, with thicknesses of 20, 30, and 50 $\upmu$m were tested. All sensors were pre-irradiated at RINSC, and subjected to proton, carbon, oxygen, iron, and gold beams. Out of 72 total tested sensors, 17 sensors exhibit behavior consistent with damage, falling into 3 categories: (1) SEB, occurring above the expected voltage threshold of $\geq$ 12 V/$\upmu$m with an associated crater, with no clear spatial dependence. (2) Permanent damage from high current, usually with an associated crater near the GR. (3) Damage from heavy ion beams or electrical effects, where increased current is observed. Within statistical uncertainty, results are independent of device type (PiN diodes or LGADs) and thickness. Of these sensors, 16 reached an E-field of 12 V/$\upmu$m.

These results are compatible with the critical field of 12 V/$\upmu$m, as found in previous studies using MIPs. Effects compatible with SEB are observed for both LGADs and PiN diodes, and for both proton and heavy ions beams, in which a crater is produced in the sensor active area. There is no clear dependence of crater position with beam type, sensor type, or thickness, given the statistics used for this study. For the heaviest ion beams used in this study, several sensors show behavior consistent with sensor damage. This is suspected to be a form of damage unique to the heavy ion beams.

\bibliographystyle{elsarticle-num}
\bibliography{bib.bib}

\section*{Acknowledgments}

The authors wish to thank their colleagues at Brookhaven National Laboratory: Dannie Steski, Tom Kubley, John Bohnenblusch, Alan Gustavsson and the entire team at the Brookhaven Tandem Van de Graaff accelerator; Chris Musso, Don Pinelli, Antonio Verderosa, Joe Pinz and Tim Kersten for sensor mounting. We also thank the staff at the Rhode Island Nuclear Science Center for their support during the preparation and execution of the neutron irradiation of the sensors. Finally we acknowledge the essential contributions by Artur Apresyan (Fermilab, US), Ryan Heller (Lawrence Berkeley National Laboratory, US), Ronald Lipton (Fermilab, US), Christopher Madrid (Texas Tech University, US) and Koji Nakamura (KEK, JP).  This material is based upon work supported by the Brookhaven National Laboratory -- Laboratory Directed Research \& Development Grants \textit{``LGAD mortality triggered by highly ionizing particles"}, \textit{``Understanding Single Event Burnout to empower future silicon detectors"}, \textit{``Impact of LGAD design and material on Single Event Burnout events"} and by the U.S. Department of Energy under grants DE-447 SC0012704, DE-SC443363, DE-SC426496 and DE-448 SC0020255. This research used resources of the Center for Functional Nanomaterials, which is a U.S. DOE Office of Science Facility, at Brookhaven National Laboratory under Contract No. DE-SC0012704.

\newpage

\begin{appendix}

\setcounter{figure}{0}

\clearpage
\section{Stopping power}
\label{appendix:StoppingPowerPlots}

\begin{figure}[htbp]
    \centering
    \includegraphics[width=.75\linewidth]{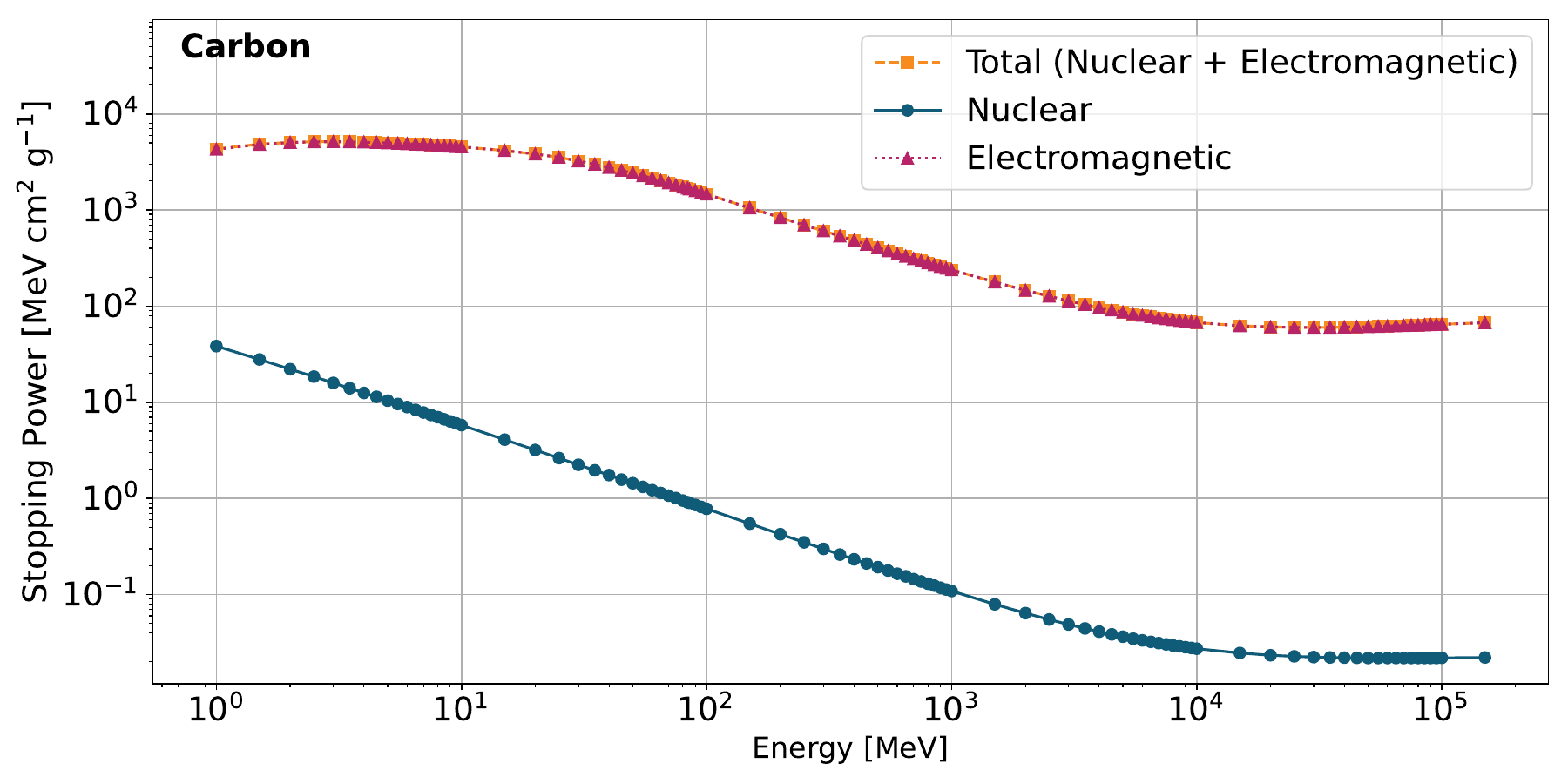}
    \caption{Comparison of electromagnetic and nuclear stopping power from SRIM simulation of carbon in Silicon, computed with \cite{srniel_calculator, srniel_calculator_nuclear}.}
    \label{fig:stoppingPower_Carbon}
\end{figure}

\begin{figure}[htbp]
    \centering
    \includegraphics[width=.75\linewidth]{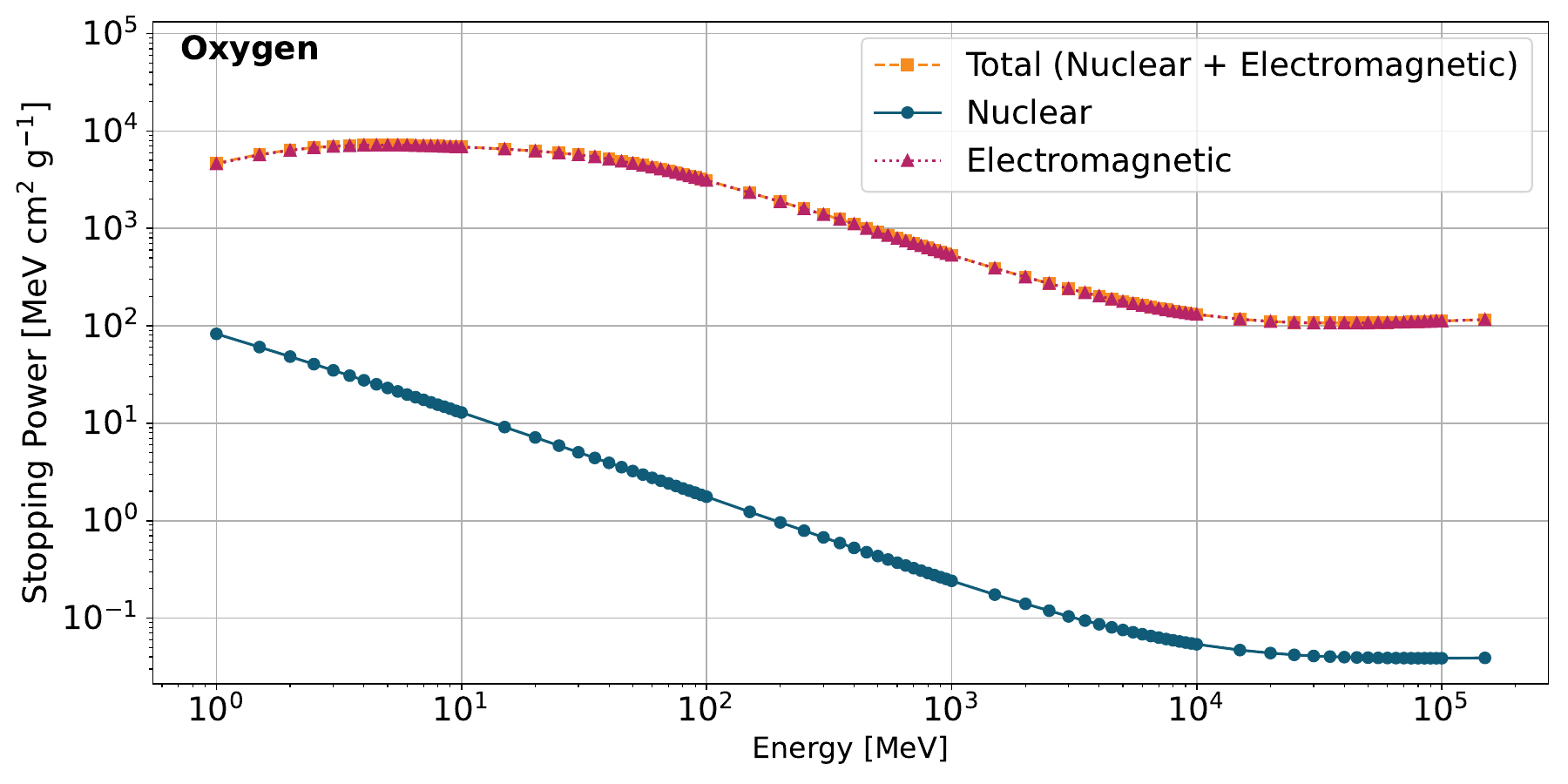}
    \caption{Comparison of electromagnetic and nuclear stopping power from SRIM simulation of oxygen in Silicon, computed with \cite{srniel_calculator, srniel_calculator_nuclear}.}
    \label{fig:stoppingPower_Oxygen}
\end{figure}

\begin{figure}[htbp]
    \centering
    \includegraphics[width=.75\linewidth]{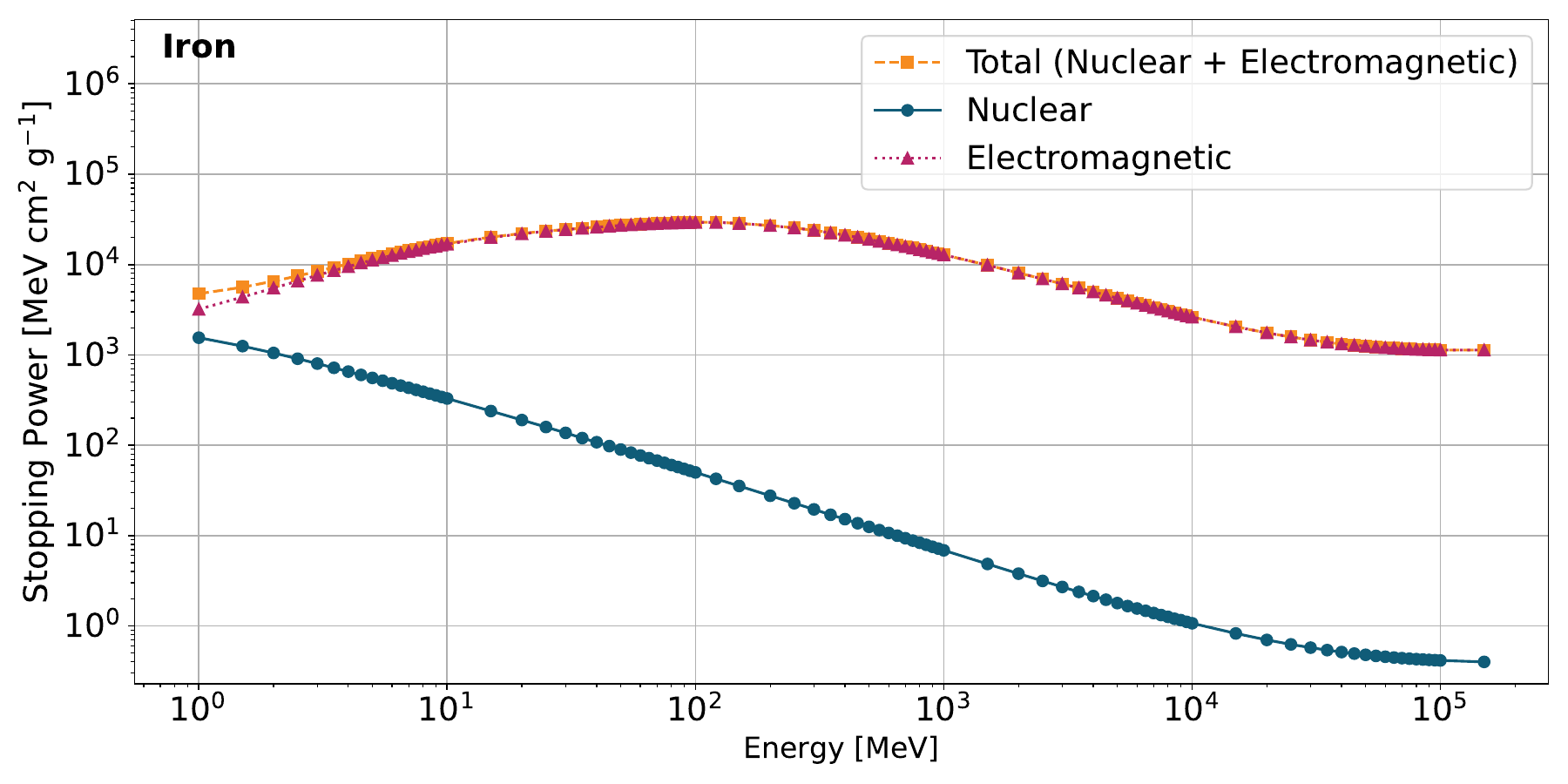}
    \caption{Comparison of electromagnetic and nuclear stopping power from SRIM simulation of iron in Silicon, computed with \cite{srniel_calculator, srniel_calculator_nuclear}.}
    \label{fig:stoppingPower_Iron}
\end{figure}

\begin{figure}[htbp]
    \centering
    \includegraphics[width=.75\linewidth]{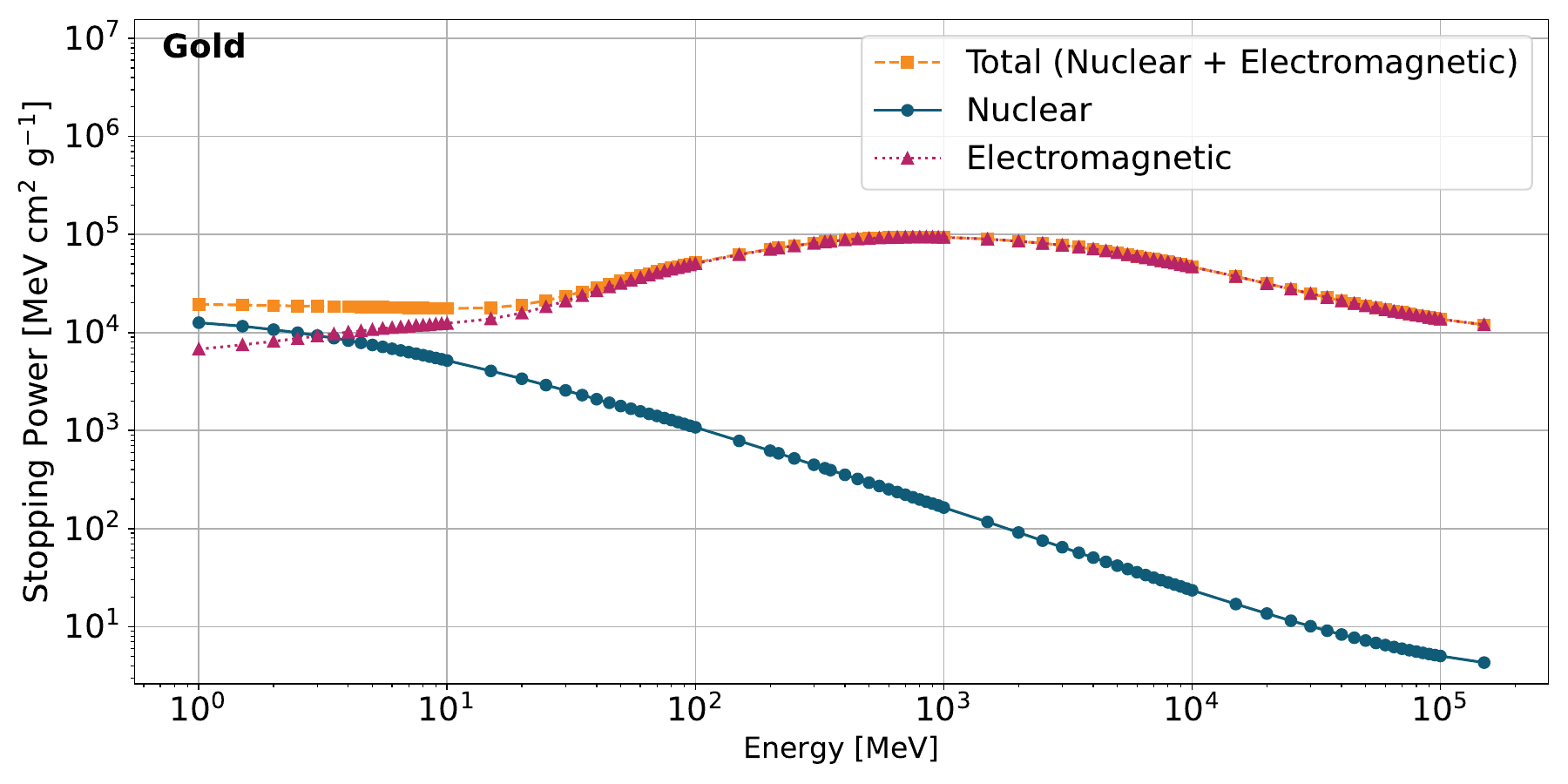}
    \caption{Comparison of electromagnetic and nuclear stopping power from SRIM simulation of gold in Silicon, computed with \cite{srniel_calculator, srniel_calculator_nuclear}.}
    \label{fig:stoppingPower_Gold}
\end{figure}

Stopping power as a function of energy is shown in Fig.~\ref{fig:stoppingPower_Carbon}, ~\ref{fig:stoppingPower_Oxygen}, ~\ref{fig:stoppingPower_Iron}, ~\ref{fig:stoppingPower_Gold}.

\setcounter{figure}{0}

\clearpage
\section{Metrology results}
\label{appendix:MetrologyResults}

\begin{figure*}[h!]
    \centering
    \begin{minipage}{0.8\textwidth}
        \begin{minipage}[b]{0.48\linewidth}
            \centering
            \includegraphics[width=\linewidth]{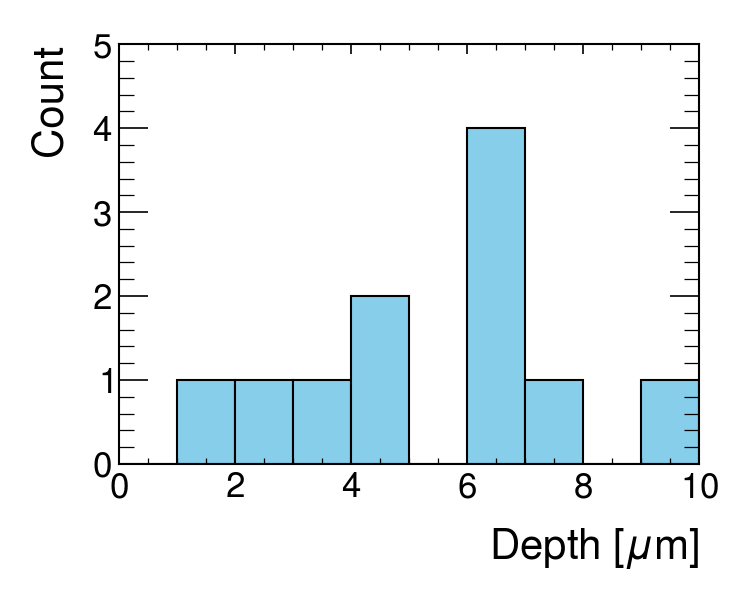}
            \captionof{figure}{Distribution of crater depths.}
            \label{fig:Crater_depths}
        \end{minipage}
        \hfill
        \begin{minipage}[b]{0.48\linewidth}
            \centering
            \includegraphics[width=\linewidth]{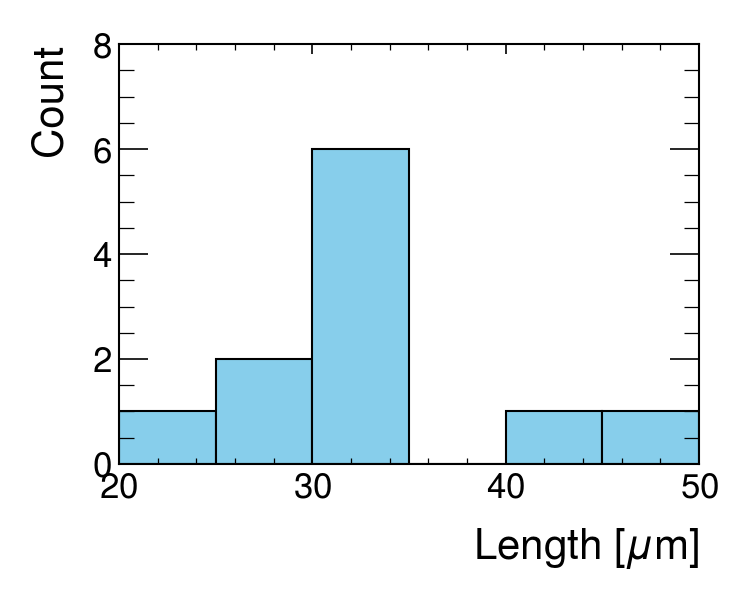}
            \captionof{figure}{Distribution of crater major axis lengths.}
            \label{fig:Crater_major}
        \end{minipage}

        \vspace{0.5em}

        \begin{minipage}[b]{0.48\linewidth}
            \centering
            \includegraphics[width=\linewidth]{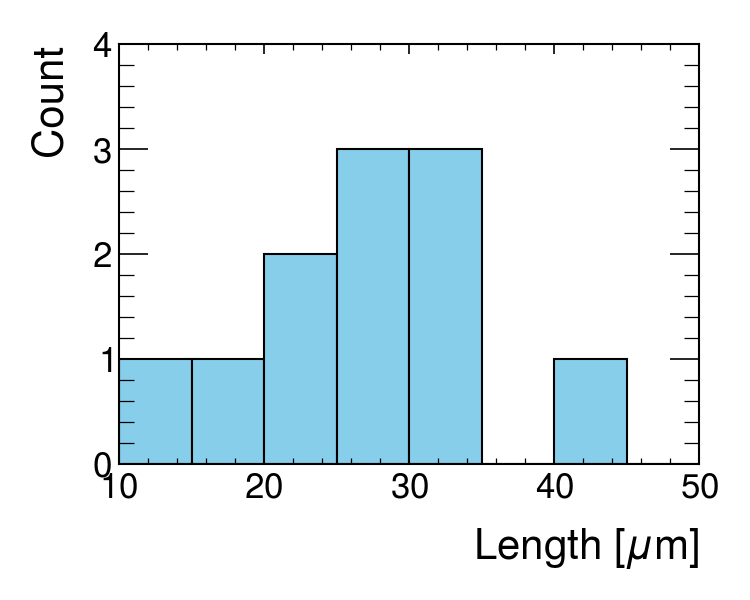}
            \captionof{figure}{Distribution of crater minor axis lengths.}
            \label{fig:Crater_minor}
        \end{minipage}
        \hfill
        \begin{minipage}[b]{0.48\linewidth}
            \centering
            \includegraphics[width=\linewidth]{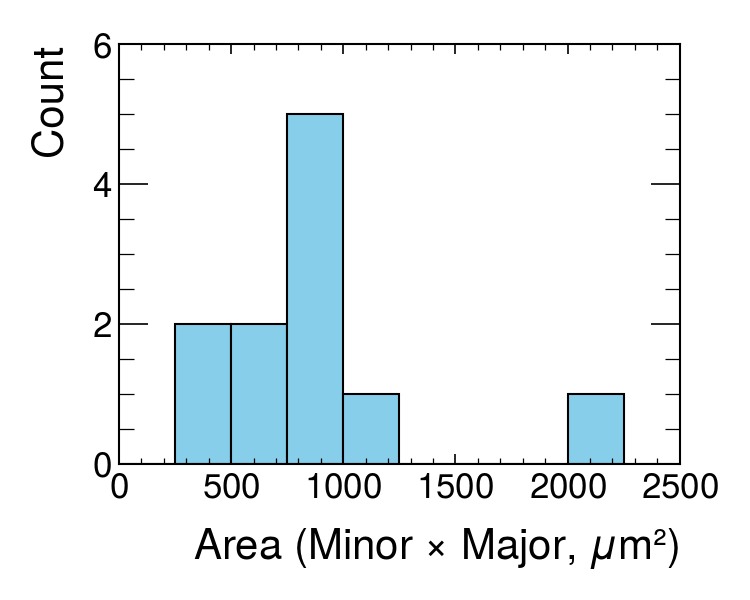}
            \captionof{figure}{Distribution of crater areas.}
            \label{fig:Crater_areas}
        \end{minipage}
    \end{minipage}
\end{figure*}

\begin{figure*}[h!]
    \centering
    \includegraphics[width=0.75\linewidth]{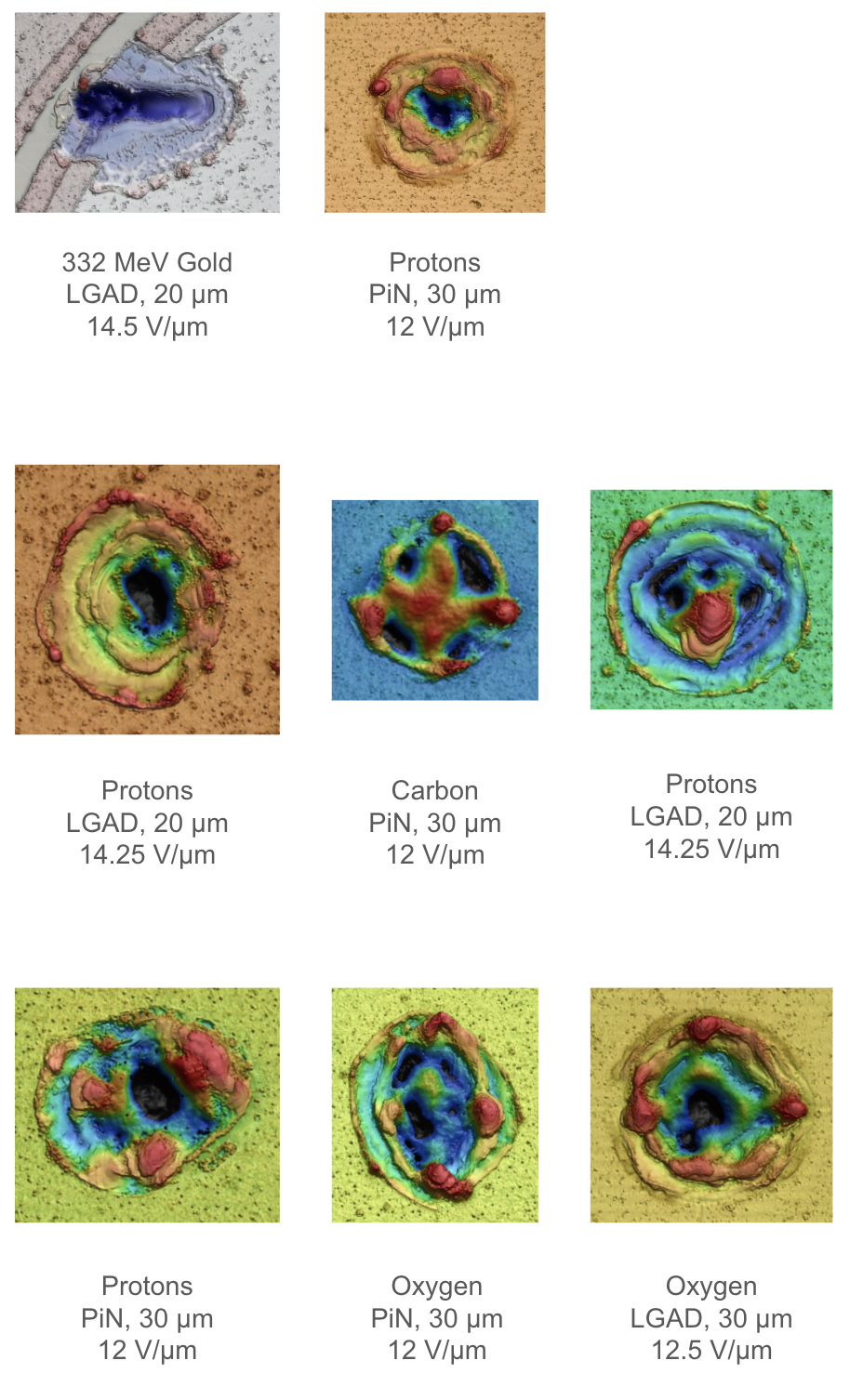}
    \caption{All craters observed for Category 1 sensors (``SEB candidates").}
    \label{fig:all_cat_1_craters}
\end{figure*}

\begin{figure*}[h!]
    \centering
    \includegraphics[width=\linewidth]{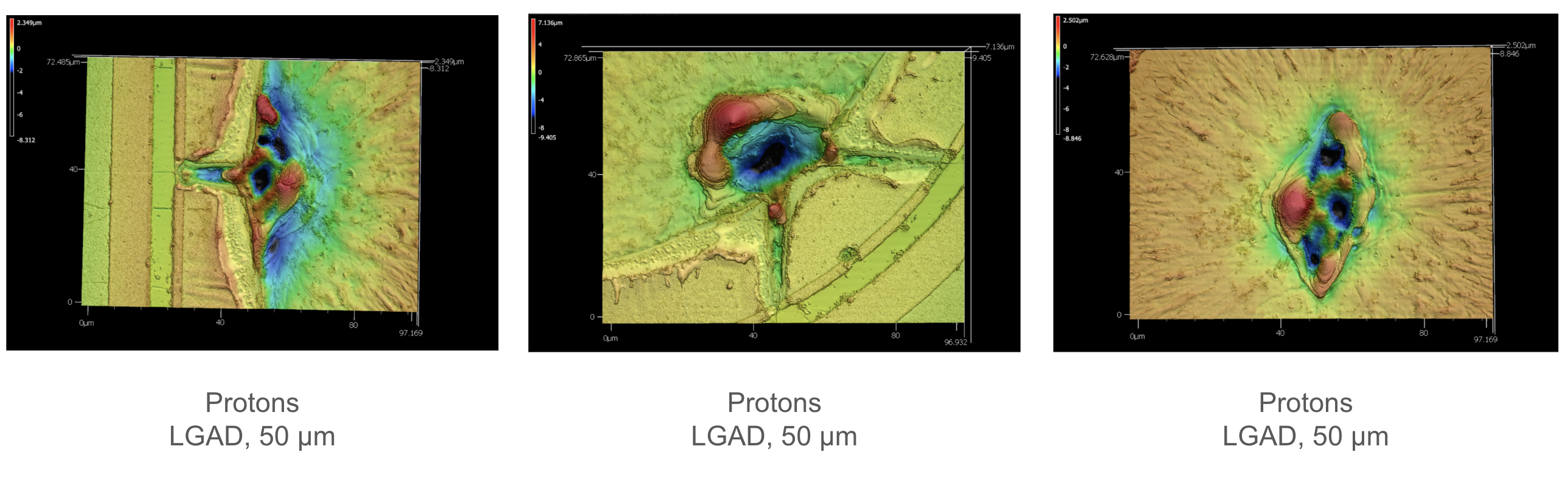}
    \caption{All craters observed for Category 2 sensors (``Damage from high current, no beam").}
    \label{fig:all_cat_2_craters}
\end{figure*}

All Category 1 craters are shown in Fig.~\ref{fig:all_cat_1_craters}, and all Category 2 craters are shown in Fig.~\ref{fig:all_cat_2_craters}. The distributions of crater dimensions, including crater depth, major axis length, minor axis length, and area are shown in Figs. \ref{fig:Crater_depths}, \ref{fig:Crater_major}, \ref{fig:Crater_minor}, \ref{fig:Crater_areas}, respectively. It can be seen in Fig.~\ref{fig:all_cat_1_craters} that most craters are circular and symmetric in nature, including some with a star-like shape. One crater with an oval-like shape was created in interaction with a 332 MeV Gold ion, and was discovered directly under a wire-bond near the guard ring. The craters in Fig.~\ref{fig:all_cat_2_craters} appear near the guard ring, indicating a metallization failure. 

\newpage

\setcounter{figure}{0}

\clearpage
\section{Estimate of number of particles}
\label{appendix:NparticlesEstimation}

Tab.~\ref{tab:Nparticles_table} shows a rough estimate of the number of ions that passed through each category 1 sensor.

\begin{table*}[htbp]
\centering

\resizebox{\textwidth}{!}{%
    \begin{tabular}{c|c|c|c|c|c|c}
    Beam type      & Sensor type  & Sensor thickness [$\mu$m] & \makecell{Estimated pre-irradiation\\fluence [$10^{15}\mathrm{n_{eq}/cm^2}$]} & Damage threshold [V/$\mu$m] & \makecell{Estimated number of ions\\below SEB threshold} & \makecell{Estimated number of ions\\at SEB threshold} \\ \hline \hline

    Protons               & PiN      & 30 & 1.56 & 12 & 8.4E9 & 12E6   \\
    Protons                & PiN      & 30 & 1.91 & 12 & 8.4E9 & 12E6    \\
    Protons                & LGAD      & 20 & 1.37 & 14.25 & 142E6 & 11.5E6    \\
    Protons                & LGAD      & 20 & 1.37 & 14.25 & 142E6 & 54E6    \\
    Carbon                & PiN      & 30 & 1.27 & 12 & 1.85E6 & 82E3   \\
    Oxygen                & PiN      & 30 & 1.27 & 12  & 1.3E6 & 252E3   \\
    Oxygen                & LGAD      & 30 & 1.27 & 12.5 & 1.1E6 & 472E3   \\
    Gold (216 MeV)        & LGAD      & 20 & 1.5 & 14.5 & 108E3 & 19E3   \\
   
    \end{tabular}
}
\caption{Rough estimate of number of particles passing through Category 1 devices at damage threshold, and at an electric field below damage threshold.\label{tab:Nparticles_table} Due to technical limitations of the facility, we could not reliably measure the instantaneous flux of the beam, and instead sampled the flux once per run. This is used to give an estimate of the number of particles which passed through each Category 1 device, assuming the sampled beam flux, which has a large inherent uncertainty.}
\end{table*}


   

\end{appendix}

\end{document}